\documentstyle[psfig,onecolumn]{mn2e}
\input{epsf}

\newif\ifAMStwofonts
\AMStwofontstrue

\newcommand{\go}{\mathrel{\raise.3ex\hbox{$>$}\mkern-14mu
             \lower0.6ex\hbox{$\sim$}}}
\newcommand{\lo}{\mathrel{\raise.3ex\hbox{$<$}\mkern-14mu
             \lower0.6ex\hbox{$\sim$}}}
\newcommand{\lp}{\left(}
\newcommand{\rp}{\right)}
\newcommand{\lb}{\left[}
\newcommand{\rb}{\right]}

\newcommand{\vecr}{\bmath r}
\newcommand{\vecB}{\bmath B}
\newcommand{\vechatB}{\hat{\bmath B}}
\newcommand{\vecmu}{\bmath \mu}
\newcommand{\vecW}{\bmath \Omega}
\newcommand{\vechatW}{\hat{\bmath \Omega}}
\newcommand{\vecE}{\bmath E}
\newcommand{\vecD}{\bmath D}

\newcommand{\vecA}{\bmath A}

\newcommand{\veck}{{\bmath k}}
\newcommand{\vechatk}{\hat{\bmath k}}
\newcommand{\hatk}{\hat{k}}

\newcommand{\vechaty}{\hat{\bmath y}}

\newcommand{\vechatz}{\hat{\bmath z}}

\newcommand{\vechatX}{\hat{\bmath X}}
\newcommand{\vechatY}{\hat{\bmath Y}}
\newcommand{\vechatZ}{\hat{\bmath Z}}

\newcommand{\dt}{{\bmath{[\bepsilon]}}}

\newcommand{\singh}{\sin\theta_B}
\newcommand{\singhsq}{\sin^2\theta_B}
\newcommand{\cosgh}{\cos\theta_B}
\newcommand{\cosghsq}{\cos^2\theta_B}
\newcommand{\ggtheta}{\lp1-\beta\cosgh\rp}
\newcommand{\ggthetas}{\lp1-\beta_s\cosgh\rp}

\newcommand{\xigh}{(\cosgh-\xi\singh)}

\newcommand{\wpl}{\omega_{\rm pl}}
\newcommand{\wc}{\omega_{\rm c}}
\newcommand{\damp}{\gamma_{\rm rad}}
\newcommand{\polarb}{\beta_{\rm pol}}
\newcommand{\PA}{\phi_{\rm PA}}
\newcommand{\intd}{{\rm d}}
\newcommand{\ri}{{r{\rm i}}}
\newcommand{\mui}{{\mu{\rm i}}}

\title[]{
Polarization Changes of Pulsars due to Wave Propagation Through Magnetospheres
} 
\author[C. Wang, D. Lai and J.L. Han]
  {Chen Wang$^{1,2}$, Dong Lai$^{2}$, JinLin Han$^{1}$ \\ 
  $^{1}$ National Astronomical Observatories, Chinese Academy of
  Sciences.  A20 Datun Road, Chaoyang District, Beijing 100012, China \\
  $^{2}$ Department of Astronomy, Cornell University, Ithaca, NY
  14853, USA \\ 
  {\rm E-mail: cwang, dong@astro.cornell.edu, hjl@nao.cas.cn}}
\date{Accepted 2009 xxx, Received 2009 xxx; in original form 2009 xxx}

\pagerange{\pageref{firstpage}--\pageref{lastpage}}
\begin{document}

\maketitle

\label{firstpage}

\begin{abstract}
We study the propagation effects of radio waves in a pulsar
magnetosphere, composed of relativistic electron-positron pair plasmas
streaming along the magnetic field lines and corotating with the
pulsar.  We critically examine the various physical effects that can
potentially influence the observed wave intensity and polarization,
including resonant cyclotron absorption, wave mode coupling due to
pulsar rotation, wave propagation through quasi-tangential regions
(where the photon ray is nearly parallel to the magnetic field) and
mode circularization due to the difference in the electron/positron
density/velocity distributions.  We numerically integrate the transfer
equations for wave polarization in the rotating magnetosphere, taking
account of all the propagation effects in a self-consistent manner.
For typical magnetospheric plasma parameters produced by pair cascade,
we find that the observed radio intensity and polarization profiles
can be strongly modified by the propagation effects. For relatively
large impact parameter (the minimum angle between the magnetic dipole
axis and the line of sight), the polarization angle profile is similar
to the prediction from the Rotating Vector Model, except for a phase
shift and an appreciable circular polarization.  For smaller impact
parameter, the linear polarization position angle may exhibit a sudden
$90^o$ jump due to the quasi-tangential propagation effect,
accompanied by complex circular polarization profile.  Some
applications of our results are discussed, including the origin of
non-gaussion pulse profiles, the relationship between the position
angle profile and circular polarization in conal-double pulsars, and
the orthogonal polarization modes.
\end{abstract}

\begin{keywords}
plasmas -- polarization -- waves -- star: magnetic fields -- pulsars:
general
\end{keywords}

\section{Introduction} \label{sec:intro}

Pulsar radio emission is likely generated within a few hundred
kilometers from the neutron star (NS) surface (e.g. Cordes 1978;
Blaskiewicz et al. 1991; Kramer et al.~1997; Kijak \& Gil 2003). A
pulsar is surrounded by a magnetosphere filled with relativistic
electron-positron pair plasmas (plus possibly a small amount of ions)
within the light cylinder. When radio waves propagate though the
magnetosphere, the total flux, polarization state and spectrum of the
emission may be modified by propagation effects.  Understanding the
property of wave propagation in pulsar magnetospheres is necessary for
the interpretation of various observations of pulsars.

Radio emission from pulsars shows strong linear polarization. For some
pulse components or even the whole pulse profiles it can be 100\%
percent polarized (e.g. Lyne \& Manchester 1988; Gould \& Lyne 1998;
Weisberg et al. 1999, 2004; Han et al. 2009).  Linear polarization
(LP) is closely related to magnetic field lines where the emission was
generated. Based on the linear polarization position angle (PA) curve
of Vela pulsar, the Rotating-Vector-Model (RVM) was suggested by
Radhakrishnan \& Cooke (1969). For some pulsars, especially the
so-called conal-double type pulsars, RVM works very well (e.g., Mitra
\& Li 2004). However, the PA curves of most pulsars are much more
complex and do not follow the simple RVM model. The deviation from the
RVM model could be caused by the intrinsic emission mechanism (e.g.,
Blaskiewicz et al. 1991), which is highly uncertain (e.g., Lyubarsky
2008), and/or the propagation effect through the pulsar magnetosphere
(see below).  Also, the PA curves or polarization observations of
individual pulses show the orthogonal polarization modes (OPM)
phenomenon, in which the polarization position angle exhibits a sudden
$\sim 90^\circ$ jumps (e.g. Manchester et al. 1975; Backer et
al. 1976; Cordes et al. 1978; Stinebring et al.~1984a, 1984b; Xilouris
et al. 1995).  It is not clear whether the OPM arises from the
emission process (e.g. Luo \& Melrose 2004) or the propagation effect
(e.g. McKinnon \& Stinebring 2000).

Another important observational feature of pulsar radio emission is
the circular polarization (CP, e.g. Rankin 1983; Radhakrishnan \&
Rankin 1990; Han et al. 1998). Significant CPs have been observed in
individual pulses of pulsars with mean values typically
20\%--30\%. Very high degrees of CP are occasionally observed from
some components of pulsar profiles (e.g. Cognard et al. 1996; Han et
al. 2009). Radhakrishnan \& Rankin (1990) identified two main types of
CP signature: antisymmetric type with sign reverse in the mid-pulse
and symmetric type without sign change over whole profile. They
concluded that the CP of the antisymmetric type is associated with the
core emission and strongly correlated with the sense of rotation of
the linear position angle. Han et al. (1998) showed that this
correlation is not kept for a larger sample, and they found that for
conal-double pulsars the sense of CP is correlated with the sense of
PA curves.

The diverse behaviours of pulsar polarization (including LP and CP)
may require more than one mechanisms for proper explanations.  First,
they may be caused by an intrinsic mechanism in the emission region
and/or process. For example, Randhakrishnan \& Rankin (1990) suggested
that geometrical effect to the pulsar beam from curvature radiation
can naturally generate antisymmetric circular polarization for the
core components. Gangadhara (1997) suggested that the observed
circular polarization could be caused by the coherent superposition of
two orthogonal modes emitted by positrons and electrons. Xu et
al. (2000) interpreted the circular polarization by the superposition
of coherent inverse Compton scattering. Kazbegi et al. (1991)
suggested that cyclotron instability may be responsible for the
circular polarization.  Also, Luo \& Melrose (2001) suggested that
circular polarization can develop by cyclotron absorption when the
distributions (especially the number densities) of the magnetospheric
electrons and positrons are different.

However, many observed characteristics of the pulsar radio emision are
most likely dictated by the wave propagation in the magnetospheric
plasma (see, e.g., Melrose 2003 and Lyubarsky 2008 for a review).  A
number of theoretical works have been devoted to study how
magnetosphere propagation influences pulsar polarization observations.
Whatever the emission mechanism, radio wave propagates in the plasma
in the form of two orthogonally polarized normal modes.  The
polarization state of the wave evolves along the ray, following the
direction of the local magnetic field, a process termed ``adiabatic
walking'' (Cheng \& Ruderman 1979).  Cheng \& Ruderman (1979)
introduced two propagation effects: the wave mode coupling effect for
pure pair plasma and the circularization effect (natural modes become
circular polarized), both of which can generate circular
polarization. Melrose (1979) and Allen \& Melrose (1982) suggested
that the separation of natural waves (because of different refractive
indices) can cause the OPM phenomenon.  Arons \& Barnard (1986)
studied the wave dispersion relation and natural modes in the
relativistic pair plasma.  Lyubaskii \& Petrova (1999) considered the
natural modes in relativistic plasma with co-rotating velocity in the
infinite magnetic field limit, and Petrova \& Lyubarskii (2000)
studied refraction and polarzation transfer in such a plasma.  Luo \&
Melrose (2001) and Fussell et al.~(2003) studied the cyclotron
absorption of radio emission within pulsar magnetospheres. Petrova
(2006) further studied the polarization transfer in pulsar
magnetosphere and considered the wave mode coupling and cyclotron
absorption effect. Johnston et al. (2005) suggested that the variation
of circular polarization of PSR B1259$-$63 during the elipse with its
main-sequence companion is related to the wave propagation effect in
the magnetosphere of the companion star.  However, none of the
previous studies have calculated the final polarization profiles with
all of these propagation effects included in a self-consistent way
within a single theoretical framework.  It is often unclear which of
the effects are most important, and if so, under what conditions.  In
this paper we attempt to combine all the propagation effects, evaluate
their relative importance, and use numerical integration along the
photon ray to study the influence of propagation effects on the final
polarization states.

This paper is organized as follows. In section 2, we present the
geometrical model for our calculation and the general wave evolution
equation in a magnetized plasma. In section 3, we give the expression
of the dielectric tensor of a relativistic pair plasma characterizing
the magnetosphere of a pulsar, and discuss the natural wave modes and
their evolution. In section 4, we study several important propagation
effects separately: cyclotron absorption, wave mode coupling,
circularization and the quasi-tangential propagation (see Wang \& Lai
2009). In section 5, we present numerical calculations of the single
photon evolution and the phase profiles of pulsar emission beam. Our
results and possible applications are presented in section 6.

\section{Geometry and General Wave Evolution Equation}

\subsection{Geometrical Model}
\begin{figure*}
\begin{tabular}{cc}
\psfig{figure=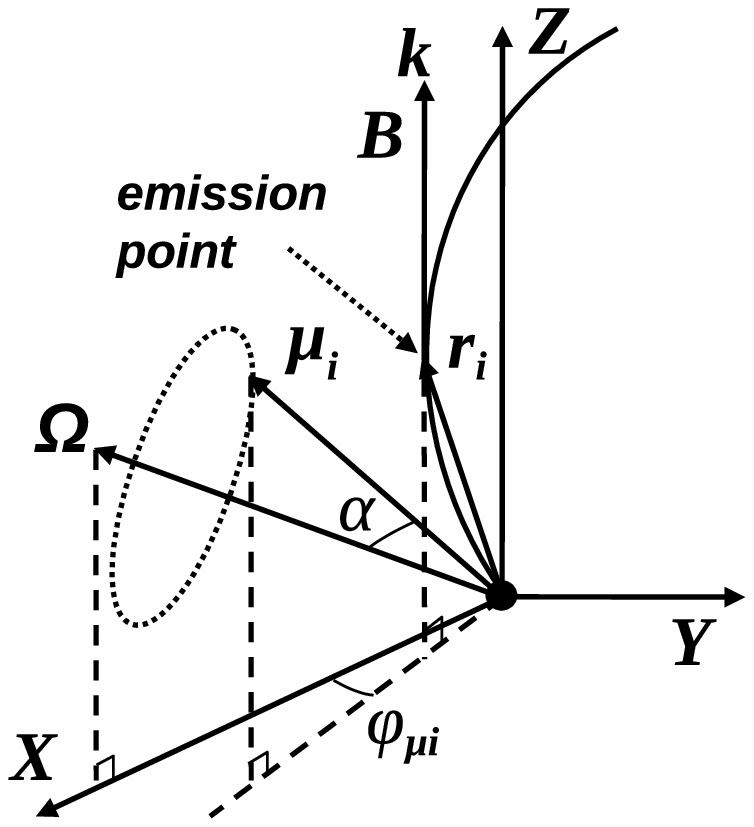,angle=0,height=8cm} &
\psfig{figure=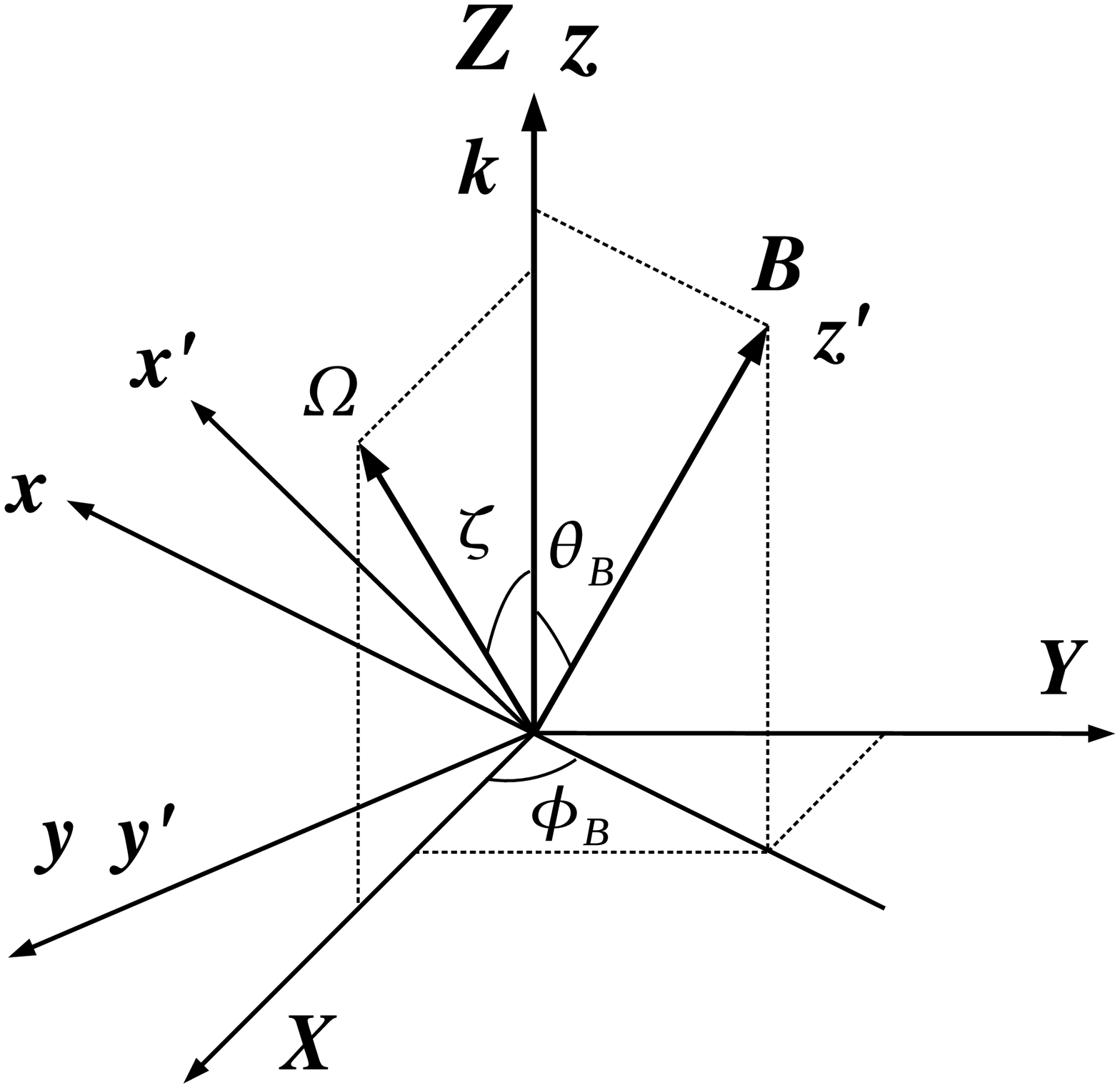,angle=0,height=8cm} \\
\end{tabular}
\caption{ Three frames used in this paper: 1) The fixed frame $XYZ$
  with $\vechatZ\parallel\vechatk$, $\vecW$ in the $XZ$-plane and
  $\vechatk \times \vechatW = \Omega \sin\zeta \vechatY$. The
  direction of $\vecB$ in this frame is ($\theta_B, \phi_B$); 2) The
  instantaneous inertial frame $xyz$ with $ \vechatz = \vechatZ$,
  $\vecB$ in the $xz$-plane and $\vechatk \times \vechatB = -
  \sin\theta_B \vechaty$; 3) The instantaneous inertial frame $x'y'z'$
  with $\vechatz'=\vechatB$, $\veck$ in the $x'z'$-plane and $\vechatk
  \times \vechatB = -\sin\theta_B\vechaty'$.\label{fig1}}
\end{figure*}

Consider a photon (radio wave) emitted at the initial position
$\vecr_{\rm i}$ at time $t_{\rm i}$ (corresponding to the pulsar
rotation phase $\Psi_{\rm i}$). Suppose the photon trajectory is a
straight line along $\veck$ (the wave vector). In a fixed $XYZ$ frame
with $\vechatZ=\vechatk$ along the line of sight and $\vecW$ (the
pulsar spin vector) in the $XZ$-plane
($\vechatZ\times\vechatW=\sin\zeta\vechatY$, here $\zeta$ is the angle
between $\veck$ and $\vecW$; see Fig.~\ref{fig1}), the photon position
after emission and the corresponding pulsar rotation phase are
\begin{equation}
\vecr = \vecr_{\rm i} + s \vechatZ,
\end{equation}
\begin{equation}
\Psi = \Psi_{\rm i} + \Omega (t-t_i) = \Psi_{\rm i} + s/r_{\rm lc}, \label{eq:Psi}
\end{equation}
where $s=c(t-t_{\rm i})$ is the distance from the emission point along
the ray, and $r_{\rm lc}=c/\Omega$ the radius of the light
cylinder. The rotating magnetic field is given by
\begin{equation}
\vecB(s)=-\nabla (\vecmu\cdot\vecr/r^3)
     =-\frac{\vecmu}{r^3}+\frac{3\vecr}{r^5}(\vecmu\cdot\vecr), \label{eq:B}
\end{equation}
with
\begin{equation}
\vecmu(s) = \mu\lb(\sin\zeta\cos\alpha-\cos\zeta\sin\alpha\cos\Psi)\vechatX -
          \sin\alpha\sin\Psi \vechatY +
      (\cos\zeta\cos\alpha + \sin\zeta\sin\alpha\cos\Psi)\vechatZ\rb,
\label{eq:mu}
\end{equation}
where $\alpha$ is the inclination angle between $\vecW$ and $\vecmu$
(see Fig.~\ref{fig1}). Note that the impact angle $\chi$, which is the
smallest angle between $\veck$ and $\vecmu$, is given by
$\chi=\zeta-\alpha$. Thus, the polar angles of $\vecmu$ in $XYZ$
frame, ($\theta_\mu$, $\phi_\mu$), are given by
\begin{equation}
\cos\theta_\mu=\cos\zeta\cos\alpha + \sin\zeta\sin\alpha\cos\Psi,\\
\tan\phi_\mu=\frac{-\sin\alpha\sin\Psi}{\sin\zeta\cos\alpha-\cos\zeta\sin\alpha\cos\Psi}.
\label{eq:theta_mu}
\end{equation}

The magnetic field at a given point along the ray
is inclined at an angle $\theta_B$ with respect to
the line of sight, and make an azimuthal angle $\phi_B$ in the
$XY$-plane such that:
\begin{equation}
\cos\theta_B(s) = \frac{B_Z}{B}, \quad
\tan\phi_B(s) = \frac{B_Y}{B_X}. \label{eq:theta_B}
\end{equation}

\subsection{Wave Evolution Equations}\label{sec:wee}

The wave equation for photon propagation takes the form
\begin{equation}
\nabla\times ({{\bmath\mu}^{-1}}\cdot\nabla\times{\vecE}) =
{\omega^2\over c^2}{\bmath\bepsilon}\cdot\vecE,
\label{eq:Maxwell}
\end{equation}
where $\vecE$ is the electric field, and $\bmath\bepsilon$,
$\bmath\mu^{-1}$ are the dielectric and inverse permeability tensors,
respectively.  The inverse permeability is very close to unity when
$B\ll B_{\rm Q}=4.414\times10^{13}$\,G (the critical QED field
strength), and we set ${\bmath\mu}$ to be unity in the remainder of
the paper. In practice, it is most convenient to calculate dielectric tensor
in the $x'y'z'$ frame (where the $z'$-axis is along $\vecB$, and
$\veck$ in the $x'z'$-plane, see Fig.~\ref{fig1}). Once $\dt_{x'y'z'}$
(the matrix representation of the dielectric tensor in the $x'y'z'$
frame) is known, we can easily obtain $\dt_{XYZ}$ in the fixed $XYZ$
frame through a coordinate transformation
\begin{equation}
\dt_{XYZ} = \bmath{M} \dt_{x'y'z'} \bmath{M}^T, \label{eq:dt_trans}
\end{equation}
where the transformation matrix $\bmath{M}$ is
\begin{equation}
\bmath{M}=
\lp\begin{array}{ccc}
-\cos\phi_B & \sin\phi_B & 0 \\
-\sin\phi_B & -\cos\phi_B & 0\\
0 & 0 & 1
\end{array}\rp
\lp\begin{array}{ccc}
\cos\theta_B & 0 & -\sin\theta_B \\
0 & 1 & 0 \\
\sin\theta_B & 0 & \cos\theta_B
\end{array}\rp =
\lp\begin{array}{ccc}
-\cos\theta_B\cos\phi_B & \sin\phi_B & \sin\theta_B\cos\phi_B \\
-\cos\theta_B\sin\phi_B & -\cos\phi_B & \sin\theta_B\sin\phi_B \\
\sin\theta_B & 0 & \cos\theta_B
\end{array}\rp\label{eq:transmatrix}
\end{equation}
and $\bmath{M}^T$ is the transpose matrix of $\bmath{M}$.

Knowing $\bmath\bepsilon$ along the trajectory, we can
use eq.~(\ref{eq:Maxwell}) to derive the wave amplitude evolution
equation. Let $\vecE=e^{ik_0s}\vecA$, where $k_0=\omega/c$. Assuming
that $|d\vecA/\intd s|\ll k_0|\vecA|$ (geometric optics
approximation), we obtain
\begin{equation}
\frac{\intd}{\intd s}\left(\begin{array}{c}A_X\\A_Y\end{array}\right)
={ik_0\over 2}\left[\begin{array}{cc}
\sigma_{XX} & \sigma_{XY} \\
\sigma_{YX} & \sigma_{YY}
\end{array} \right]
\left(\begin{array}{c}A_X\\A_Y\end{array}\right),
\label{eq:evol}
\end{equation}
where
\begin{eqnarray}
\sigma_{XX} &=& \epsilon_{XX}-1,\nonumber \\
\sigma_{XY} &=& \epsilon_{XY},\nonumber \\
\sigma_{YX} &=& \epsilon_{YX},\nonumber \\
\sigma_{YY} &=& \epsilon_{YY}-1. \label{eq:sigma_evol}
\end{eqnarray}
The wave evolution equation~(\ref{eq:evol}) can be used to study the evolution of EM wave
amplitude across the pulsar magnetosphere.

We can also follow the evolution of the four Stokes parameters instead
of the evolution of the wave amplitudes. The four Stokes parameters
are defined by (in the fixed $XYZ$ frame)
\begin{eqnarray}
I &=& A_XA_X^\ast+A_YA_Y^\ast,\nonumber\\
Q &=& A_XA_X^\ast-A_YA_Y^\ast,\nonumber\\
U &=& A_XA_Y^\ast+A_YA_X^\ast,\nonumber\\
V &=&-i(A_XA_Y^\ast-A_YA_X^\ast).
\end{eqnarray}
Combining with eq.~(\ref{eq:evol}), we obtain the evolution equations
for the Stokes parameters:
\begin{eqnarray}
\frac{\intd I}{\intd s}&=& -k_0\frac{\sigma_{XX,i}+\sigma_{YY,i}}{2}I
                 -k_0\frac{\sigma_{XX,i}-\sigma_{YY,i}}{2}Q
                 -k_0\frac{\sigma_{XY,i}+\sigma_{YX,i}}{2}U
                 -k_0\frac{\sigma_{XY,r}-\sigma_{YX,r}}{2}V, \nonumber\\
\frac{\intd Q}{\intd s}&=& -k_0\frac{\sigma_{XX,i}-\sigma_{YY,i}}{2}I
                 -k_0\frac{\sigma_{XX,i}+\sigma_{YY,i}}{2}Q
                 -k_0\frac{\sigma_{XY,i}-\sigma_{YX,i}}{2}U
                 +k_0\frac{\sigma_{XY,r}+\sigma_{YX,r}}{2}V, \nonumber\\
\frac{\intd U}{\intd s}&=& -k_0\frac{\sigma_{XY,i}+\sigma_{YX,i}}{2}I
                 +k_0\frac{\sigma_{XY,i}-\sigma_{YX,i}}{2}Q
                 -k_0\frac{\sigma_{XX,i}+\sigma_{YY,i}}{2}U
                 -k_0\frac{\sigma_{XX,r}-\sigma_{YY,r}}{2}V, \nonumber\\
\frac{\intd V}{\intd s}&=&  k_0\frac{\sigma_{XY,r}-\sigma_{YX,r}}{2}I
                 -k_0\frac{\sigma_{XY,r}+\sigma_{YX,r}}{2}Q
                 +k_0\frac{\sigma_{XX,r}-\sigma_{YY,r}}{2}U
                 -k_0\frac{\sigma_{XX,i}+\sigma_{YY,i}}{2}V. 
\label{eq:IQUV_evol}
\end{eqnarray}
Here the subscript ``i'' and ``r'' correspond to the real and imaginary
part of each element.

If we know the dielectric tensor along the ray, we
can integrate eq.~(\ref{eq:evol}) from the emission point in the inner
magnetosphere to large distance where the plasma no longer affect the radiation
(both intensity and polarization). We will calculate the dielectric tensor
of a relativistic streaming pair plasma in the next section.

\section{Wave Modes and Propagation in a Streaming Plasma}

The magnetospheres of pulsars consist of relativistic
electron-positron pair plasma streaming along magnetic field lines.
The Lorentz factor $\gamma$ of the streaming motion and the plasma
density $N$ are uncertain.  For the open field line region of radio
pulsars, pair cascade simulations generally give $\gamma\sim
10^2-10^4$ and $\eta\equiv N/N_{\rm GJ}\sim 10^2-10^5$ (e.g.,
Daugherty \& Harding 1982; Hibschman \& Arons 2001; Medin \& Lai
2009), while recent theoretical works suggest that the corona of
magnetars consist of pair plasma with $\gamma$ up to $10^3$ and
$\eta\sim 2\times 10^3(R_\ast/r)$ (where $R_\ast$ is the stellar
radius; Thompson et al.~2002; Beloborodov \& Thompson 2007). Here
$N_{\rm GJ} =(\Omega B)/(2\pi ec)$ is the Goldreich-Julian density.
In this paper, we choose plasma density $\eta=N/N_{\rm GJ}$ to be in
the range of 100~--~1000, and the Lorentz factor $\gamma$ of the
streaming motion to be 100~--~1000.  We also consider a small
asymmetry between positrons and electrons, i.e. $\Delta N/N \neq0$ and
$\Delta\gamma/\gamma \neq0$, where $\Delta N$, $\Delta\gamma$ are the
differences in the number densities and Lorentz factors between
electrons and positrons.

\subsection{Dielectric tensor}

The dielectric tensor $\bmath\bepsilon$ in the $x'y'z'$ frame (with
$\vechatz'=\vechatB$, $\veck$ in the $x'z'$-plane and
$\vechatk\times\vechatB=-\sin\theta_B\vechaty'$; see Fig.~\ref{fig1})
can be written as [see eqs.~(2.11)~--~(2.13) and (2.19) of Wang \& Lai
  2007]:
\begin{equation}
\dt_{x'y'z'}
= \lb \begin{array}{ccc}
\epsilon_{x'x'}  & \epsilon_{x'y'} & \epsilon_{x'z'} \\
\epsilon_{y'x'}  & \epsilon_{y'y'} & \epsilon_{y'z'} \\
\epsilon_{z'x'}  & \epsilon_{z'y'} & \epsilon_{z'z'}
\end{array} \rb, \label{eq:dt}
\end{equation}
where
\begin{eqnarray}
\epsilon_{x'x'}  &=& 1 +\sum_s\int f_{s,11} f_s(\gamma_s)d\gamma_s , \nonumber\\
\epsilon_{x'y'}  &=& -\epsilon_{y'x'}=i\sum_s\int f_{s,12} f_s(\gamma_s)d\gamma_s , \nonumber\\
\epsilon_{x'z'}  &=& \epsilon_{z'x'}=-i\sum_s\int \xi_s f_{s,11} f_s(\gamma_s)d\gamma_s , \nonumber\\
\epsilon_{y'z'}  &=& \epsilon_{z'y'}=\sum_s\int \xi_s f_{s,12} f_s(\gamma_s)d\gamma_s , \nonumber\\
\epsilon_{z'z'}  &=& 1+\sum_s\int\lp f_{s,\eta}+\xi_s^2 f_{s,11} \rp
           f_s(\gamma_s)d\gamma_s. \label{eq:items_dt}
\end{eqnarray}
with
\begin{eqnarray}
f_{s,11} &=& -\frac{v_s\gamma_s^{-1}(1+i\damp)}{(1+i\damp)^2-u_s\gamma_s^{-2}\ggthetas^{-2}} , \nonumber\\
f_{s,12} &=& -\frac{{\rm sign}\lp q_s\rp u_s^{1/2}v_s\gamma_s^{-2}\ggthetas^{-1}}
                 {(1+i\damp)^2- u_s\gamma_s^{-2}\ggthetas^{-2}}, \nonumber\\
f_{s,\eta} &=& -\frac{v_s}{(1+i\damp)\gamma_s^3 \ggthetas^2}, \nonumber \\
\xi_s    &=& \frac{n\beta_s\singh}{1-n\beta_s\cosgh}. \label{eq:f}
\end{eqnarray}
Here the subscript ``$s$'' specifies different species (``e'' is for
electron and ``p'' for positron), and $\beta_s$, $\gamma_s$ and
$f_s(\gamma_s)$ are the velocity (divided by $c$), Lorentz factor and
its distribution function. The dimenssionless parameters $u_s$, $v_s$
are
\begin{equation}
u=\frac{\omega_{\rm c}^2}{\omega^2}, \quad u_s=u,
\end{equation}
\begin{equation}
v=\frac{\wpl^2}{\omega^2}, \quad v_s=\frac{N_s}{N}v.
\end{equation}
Here $N_s$ is the number density of particles, $N=N_p+N_e$, $\wc$ and
$\wpl$ are the cyclotron and plasma frequencies, which are given by
\begin{equation}
\nu_{\rm c} = \frac{\omega_{\rm c}}{2\pi} = \frac{1}{2\pi}\frac{eB}{m_ec}
= 2.795\times10^9\,B_{12}\mbox{ GHz} \label{eq:nuc}
\end{equation}
\begin{equation}
\nu_{\rm pl} = \frac{\wpl}{2\pi} =\frac{1}{2\pi}\sqrt{\frac{4\pi N e^2}{m_e}}
= 8.960\times10^3 N^{1/2}\,{\rm Hz}
= 2.370\,\eta^{1/2}B_{12}^{1/2}P_{\rm 1s}^{-1/2}\,{\rm GHz}, \label{eq:nu_pc}
\end{equation}
where the magnetic field $B_{12}=B/(10^{12}\,{\rm G})$, the pulsar spin
period $P_{\rm 1s}=P/(1\,{\rm s})$, and the dimensionless density
$\eta=N/N_{\rm GJ}$ is measured in units of the Goldreich-Julian
density, $N_{\rm GJ}= \Omega B/(2\pi ec)\simeq 7.0\times
10^{10}B_{12}P_{\rm 1s}$ cm$^{-3}$.  The refractive index,
$n=ck/\omega$, is generally very close to unity, so we always set
$n\simeq1$ here. The radiative damping
\begin{equation}
\damp=\frac{4e^2\wc}{3mc^3} \label{eq:gg_re}
\end{equation}
is important only near the cyclotron resonance [where $\gamma_s\omega
  (1-n\beta\cosgh) \simeq \omega_{\rm c}$] and can be neglected at
other places. The function ${\rm sign}(q_s)$ equals to $-1$ for
electrons and $1$ for positrons.

In this paper we focus on cold streaming plasmas, which means that
both electrons and positrons in the streaming plasma have single
$\gamma_s$ or $f_s(\gamma_s)=\delta(\gamma_s-\gamma_{s,0})$. Thus, we
need not to integrate across $\gamma_s$ when calculating each element
of the dielectric tensor in eq.~(\ref{eq:items_dt}).

When we consider the region $r\ll r_{\rm cyc}$ (the cyclotron
resonance radius), we can take the infinite magnetic field limit, and
the damping term can be neglected.  In this case the dielectric tensor
becomes very simple (e.g., Arons \& Banard 1986)
\begin{equation}
\dt_{x'y'z'}
= \lb \begin{array}{ccl}
1 & 0 & 0 \\
0 & 1 & 0  \\
0 & 0 & 1+f_\eta
\end{array} \rb. \label{eq:dt_Binf}
\end{equation}
with $f_\eta=-v\gamma^{-3}(1-\beta\cos\theta_B)^{-2}$.

\subsection{Wave evolution equation for single-$\gamma$ plasma}

In this subsection we consider the polarization evolution equation for
single $\gamma$ plasma, i.e. all electrons (positrons) have the same
$\gamma_e$($\gamma_p$). We assume that there is a small asymmetry
between electrons and positrons in $N_s$ or $\gamma_s$: $\Delta N/N\ll
1$, where $N=N_p+N_e$, $\Delta N=N_p-N_e$ (usually $\Delta N/N$ is the
reciprocal of the multiplicity of the cascade), and/or
$\Delta\gamma/\gamma\ll1$ [where $\gamma=(\gamma_p+\gamma_e)/2$,
  $\Delta\gamma=\gamma_p-\gamma_e$].  In this case the final matrix
elements in the wave evolution equation~(\ref{eq:evol}) are
\begin{eqnarray}
\sigma_{XX}&=& F_{11}(1+f_\theta\cos^2\phi_B) +F_\eta\singhsq\cos^2\phi_B
           \simeq F_{11}+F_\eta\singhsq\cos^2\phi_B 
,\nonumber\\
\sigma_{XY}&=& F_{11}f_\theta +F_\eta\singhsq\sin\phi_B\cos\phi_B-iF_{12}
           \simeq F_\eta\singhsq\sin\phi_B\cos\phi_B-iF_{12}
,\nonumber\\
\sigma_{YX}&=& F_{11}f_\theta +F_\eta\singhsq\sin\phi_B\cos\phi_B+iF_{12}
           \simeq F_\eta\singhsq\sin\phi_B\cos\phi_B+iF_{12} 
,\nonumber\\
\sigma_{YY}&=& F_{11}(1+f_\theta\sin^2\phi_B)+F_\eta\singhsq\sin^2\phi_B
           \simeq F_{11}+F_\eta\singhsq\sin^2\phi_B
,\label{eq:sigma_singlegg}
\end{eqnarray}
with
\begin{eqnarray}
F_{11}&=&\sum_sf_{s,11}\simeq -\frac{v\gamma^{-1}}{1+2i\damp
                       -u\gamma^{-2}(1-\beta\cosgh)^{-2}}
       =F_{11,r}+iF_{11,i}, \nonumber\\
F_{12}&=&\sum_sf_{s,12}= -\sum_s\frac{{\rm sign}(q_s) v_s\gamma_s^{-}
                       u_s^{1/2}\gamma_s^{-}\ggthetas^{-1}}
                 {1+2i\damp- u_s\gamma_s^{-2}\ggthetas^{-2}}
         =F_{12,r}+iF_{12,i}, \nonumber\\
F_\eta&=&\sum_sf_{s,\eta}\simeq -v\gamma^{-3}(1-\beta\cosgh)^{-2},\nonumber\\
f_\theta &=&(\cosgh-\xi\singh)^2-1
         \simeq -\frac{4\theta_B^2\gamma^2}{(1+\theta_B^2\gamma^2)^2}.
\label{eq:dt_F_singlegg}
\end{eqnarray}
In the deriving of eq.~(\ref{eq:sigma_singlegg}), we have assumed
$\theta_B\gamma\gg 1$ (which is valid for most places), so that
$f_\theta\simeq 0$.

Using eqs.~(\ref{eq:IQUV_evol}) and (\ref{eq:sigma_singlegg}), we can
write the evolution equation of the four stokes parameters as
\begin{eqnarray}
\frac{\intd I}{\intd s}&=&-k_0F_{11,i}I-k_0F_{12,i}V,\nonumber\\
\frac{\intd Q}{\intd s}&=&-k_0F_{11,i}Q+k_0F_{12,r}U+\frac{k_0}{2}F_\eta\singhsq\sin2\phi_B V,\nonumber\\
\frac{\intd U}{\intd s}&=&-k_0F_{12,r}Q-k_0F_{11,i}U-\frac{k_0}{2}F_\eta\singhsq\cos2\phi_B V,\nonumber\\
\frac{\intd V}{\intd s}&=&-k_0F_{12,i}I-\frac{k_0}{2}F_\eta\singhsq(Q\sin2\phi_B-U\cos2\phi_B)
              -k_0F_{11,i}V. \label{eq:IQUV_evol_singlegg}
\end{eqnarray}
Here $k_0=c/\omega$, the subscript ``r'' and ``i'' specify the real
and imaginary parts.  Equation~(\ref{eq:IQUV_evol_singlegg}) is useful
for understanding the different kinds of propagation effects on the
polarization evolution (see section 4).

\subsection{Wave modes}

Using the electric displacement $\vecD={\bmath\epsilon}\cdot\vecE$ in
the Maxwell equations, we obtain the equation for plane waves with
$\vecE\propto e^{i(\veck\cdot\vecr-\omega t)}$
\begin{equation}
[\epsilon_{ij}+n^2(\hatk_i\hatk_j-\delta_{ij})]E_j=0,
\end{equation}
where $n=ck/\omega$ is the refractive index and $\vechatk=\veck/k$.
In the coordinate system $xyz$ with $\veck$ along the $z$-axis and
$\vecB$ in the $xz$-plane (see Fig.~1), we project the above equation
in the $xy$-plane and obtain
\begin{equation}
\lp\begin{array}{cc}
\eta_{xx}-n^2 & \eta_{xy} \\
\eta_{yx} & \eta_{yy}-n^2 \end{array}\rp
\lp\begin{array}{c}
E_x\\Ey\end{array}\rp
=0,\label{eq:etaE}
\end{equation}
where
\begin{eqnarray}
\eta_{xx}&=&\epsilon_{xx}-\epsilon_{xz}\epsilon_{zx}/\epsilon_{zz},\nonumber\\
\eta_{xy}&=&\epsilon_{xy}-\epsilon_{xz}\epsilon_{zy}/\epsilon_{zz},\nonumber\\
\eta_{yx}&=&\epsilon_{yx}-\epsilon_{yz}\epsilon_{zx}/\epsilon_{zz},\nonumber\\
\eta_{yy}&=&\epsilon_{yy}-\epsilon_{yz}\epsilon_{zy}/\epsilon_{zz}.
\label{eq:eta}
\end{eqnarray}
From eq.~(\ref{eq:etaE}), we obtain two eigenmodes, to be labeled as
the plus ``+'' mode and minus ``$-$'' mode. The refractive indices of the
two modes are given by
\begin{equation}
n_\pm^2=\frac{(\eta_{xx}+\eta_{yy})
\pm\sqrt{(\eta_{xx}-\eta_{yy})^2+4\eta_{xy}\eta_{yx}}}{2}.
\label{eq:n}
\end{equation}
We write the mode polarization vector as $\vecE_{\pm}=\vecE_{\pm
  T}+\vecE_{\pm z}\vechatz$ in the $xyz$-frame, with the transverse
part given by
\begin{equation}
\vecE_{\pm T}=\frac{1}{(1+K_{\pm}^2)^{1/2}}(K_{\pm},1),
\label{eq:eigenvector}
\end{equation}
where 
\begin{equation}
K_\pm=\lp\frac{E_x}{E_y}\rp_\pm = -\frac{\eta_{yy}-n_\pm^2}{\eta_{yx}}
    =\frac{(\eta_{xx}-\eta_{yy})\pm\sqrt{(\eta_{xx}-\eta_{yy})^2+4\eta_{xy}\eta_{yx}}}{2\eta_{yx}},
\label{eq:K}
\end{equation}
describes the polarization state of the two eigenmodes. 

From the dielectric tensor of relativistic streaming pair plasma given
by eqs.~(\ref{eq:dt})--(\ref{eq:f}), we obtain the tensor components
in the $xyz$ coordinate system:
\begin{eqnarray}
\epsilon_{xx} &=& \epsilon_{x'x'}\cosghsq+\epsilon_{z'z'}\singhsq
            -(\epsilon_{x'z'}+\epsilon_{z'x'})\singh\cosgh, \nonumber \\
\epsilon_{yy} &=& \epsilon_{y'y'}, \nonumber\\
\epsilon_{zz} &=& \epsilon_{x'x'}\singhsq+\epsilon_{z'z'}\cosghsq
            +(\epsilon_{x'z'}+\epsilon_{z'x'})\singh\cosgh, \nonumber \\
\epsilon_{xy} &=& -\epsilon_{yx} = \epsilon_{x'y'}\cosgh-\epsilon_{z'y'}\singh, \nonumber \\
\epsilon_{xz} &=& \epsilon_{zx}~~=  (\epsilon_{x'x'}-\epsilon_{z'z'})\singh\cosgh
            +\epsilon_{x'z'})(\cosghsq-\singhsq),\nonumber\\
\epsilon_{yz} &=& -\epsilon_{zy} = \epsilon_{y'x'}\singh+\epsilon_{y'z'}\cosgh, \label{eq:dt_xyz}
\end{eqnarray}
Combining the above equations with eq.~(\ref{eq:etaE}), we find that
$\eta_{xy}=-\eta_{yx}$, and $\eta_{yx}$ is almost purely imaginary
(except very close to cyclotron resonance). We define the polarization
parameter, $\polarb$, as
\begin{equation}
\polarb = -i\frac{\eta_{xx}-\eta_{yy}}{2\eta_{yx}}
   \simeq -i\frac{\epsilon_{y'y'}-\epsilon_{x'x'}\cos^2\theta_B-\epsilon_{z'z'}\sin^2\theta_B
  +(\epsilon_{x'z'}+\epsilon_{z'x'})\sin\theta_B\cos\theta_B}
        {2(\epsilon_{y'x'}\cos\theta_B-\epsilon_{y'z'}\sin\theta_B)}
\label{eq:polarb1}
\end{equation}
Then eq.~(\ref{eq:K}) can be written as
\begin{equation}
iK_{\pm} = \polarb\mp{\rm sign}(\eta_{yx,i})\sqrt{\polarb^2+1}.
 \label{eq:iK}
\end{equation}
Here sign($\eta_{yx,i}$) means the sign of the imaginary part of
$\eta_{yx}$.  Obviously, when $|\polarb|\gg 1$, the two eigenmodes are
linear polarized, while for $|\polarb| =0 $ the two modes are
circular-polarized.

Consider a cold pair plasma with $\Delta N=N_p-N_e\ll N$ and
$\Delta\gamma = \gamma_p - \gamma_e \ll \gamma$.  When the
Lorentz-shifted frequency, $\gamma\omega\ggtheta$, is much less than
the cyclotron frequency $\wc$, i.e. for $r\ll r_{\rm cyc}$ or $\lambda
= \wc/[\gamma\omega\ggtheta]= u^{1/2}\gamma^{-1}\ggtheta^{-1}\gg 1$,
we have
\begin{equation}
\polarb \simeq \frac{-\lambda\theta_B^2\gamma^2(1+\theta_B^2\gamma^2)^{-1}}
{(1-\theta_B^2\gamma^2)\Delta N/N-\Delta\gamma/\gamma}.
\label{eq:polarb_beforecyc}
\end{equation}
Here we assume $\theta_B\ll 1$, so that $\lambda \simeq
4u^{1/2}\gamma(1+\theta_B^2\gamma^2)^{-1}$. After the photon passes
through the cyclotron resonance, $r\gg r_{\rm cyc}$ or $\lambda\ll 1$,
the polarization paramerter is given by
\begin{equation}
\polarb \simeq \frac{\lambda \theta_B^2\gamma^2}
{\theta_B^2\gamma^2(3-\theta_B^2\gamma^2)\Delta\gamma/\gamma - (1-\theta_B^4\gamma^4)\Delta N/N}.
\label{eq:polarb_aftercyc}
\end{equation}
These expressions are useful for understanding the effect of mode circularization
(section 4.3).

\subsection{Evolution of Mode Amplitude}

In the $xyz$ frame [with $\vechatz = \vechatk$, $\vechatB = (-\sin\theta_B,
0, \cos\theta_B)$ in this frame], we know there are two wave modes:
``+'' mode and ``$-$'' mode. It is convenient to introduce a mixing
angle, $\theta_m$, via $\tan\theta_m=1/(iK_+)$, so that
\begin{equation}
\tan 2\theta_m = \polarb^{-1}.\label{eq:theta_m}
\end{equation}
In the $xyz$ frame, the transverse components of the mode eigenvectors are
\begin{equation}
\vecE_+=\left(\begin{array}{c}i\cos\theta_m\\\sin\theta_m\end{array}\right), 
\quad 
\vecE_-=\left(\begin{array}{c}-i\sin\theta_m\\\cos\theta_m\end{array}\right), 
\end{equation}
In the fixed $XYZ$ frame (see Fig.~\ref{fig1}), they become
\begin{equation}
\vecE_+=\left(\begin{array}{c}i\cos\theta_m\cos\phi_B-\sin\theta_m\sin\phi_B\\
                              i\cos\theta_m\sin\phi_B+\sin\theta_m\cos\phi_B\end{array}\right), 
\quad 
\vecE_-=\left(\begin{array}{c}-i\sin\theta_m\cos\phi_B-\cos\theta_m\sin\phi_B\\
                              -i\sin\theta_m\sin\phi_B+\cos\theta_m\cos\phi_B\end{array}\right), 
\end{equation}
The general wave amplitude can be written as
\begin{equation}
\left(\begin{array}{c}A_{X}\\A_{Y}\end{array}\right)
=A_{+}\vecE_++
 A_{-}\vecE_-.\\ 
\end{equation}
Substitute this into the wave equation, we obtain the mode amplitude
evolution equation:
\begin{equation}
i{\intd\over \intd s}\left(\begin{array}{c}A_+\\A_-\end{array}\right)
=\left[\begin{array}{cc}
-\Delta k/2+\phi_B'\sin2\theta_m & i\theta_m'+\phi_B'\cos2\theta_m \\
-i\theta_m'+\phi_B'\cos2\theta_m & \Delta k/2-\phi_B'\sin2\theta_m
\end{array} \right]
\left(\begin{array}{c}A_+\\A_-\end{array}\right)
\label{eq:me}
\end{equation}
where the superscript ($'$) specifies $\intd/\intd s$, $\Delta k=
k_+-k_-=\Delta n\omega/c$, and we have subtracted a non-essential
unity matrix from the above.  This equation generalizes the special
cases (where only $\theta_m$ or $\phi_B$ varies) studied in Lai \& Ho
(2002,2003) and van Adelsberg \& Lai (2006), and it is useful for
understanding the effect of mode coupling (section 4.2).

\section{Some Important Propagation Effects}

With the equations derived in previous sections, we can now identify
several key physical effects relevant for the evolution of wave
polarization. We consider the ``weak dispersion''
region where the wave frequency is much larger than the plasma
frequency in the plasma rest frame and the refractive indices of the
two natural wave modes are very close to unity. So we do not
discuss the refraction effect here. The detail discussion about
refraction effect can be found in Barnard \& Arons (1986).

\subsection{Cyclotron Resonance/Absorption}
\label{sec:cyc}

Cyclotron resonance occurs when the wave frequency in the
electron/positron rest frame is close to the cyclotron frequency:
\begin{equation}
\tilde{\omega}=\gamma\omega(1-\beta\cosgh)
=\wc=\frac{eB}{mc}.\label{eq:cycres}
\end{equation}
The eigenmodes at cyclotron resonance point are always two circular
polarized modes (marked as ``$+$
'' for the left-handed circular
polarized mode and ``$-$'' for the right-handed one). Since the
electrons and positrons have different directions of gyration (one is
right-handed, the other one is left-handed), the right-handed circular
polarized mode is absorbed by electrons while the left-handed circular
polarized mode absorbed by positrons.  For right-handed circular
polarized mode, the scattering cross-section by electrons in the
electron rest frame (the physical quantities in the rest frame are
marked by `` $^\sim$ '') is
\begin{equation}
\tilde{\sigma}_-\simeq
(2\pi)^2\frac{e^2}{mc}\delta(\tilde{\omega}-\wc)\label{eq:cross+}.
\end{equation}
The opitical depth of this mode in the rest frame is
\begin{equation}
\tilde{\tau}_-=\int \tilde{N_e}\tilde{\sigma}\intd \tilde{s}
\end{equation}
Since the optical depth is Lorentz invariant, and
\begin{equation}
\tilde{N}_e=\gamma^{-1}N_e, \quad \intd \tilde{s}=\gamma_e(1-\beta_e\cosgh) \intd s,
\end{equation}
the optical depth in the ``lab'' frame is
\begin{equation}
\tau_-=\tilde{\tau}_-=\int N_e\tilde{\sigma}(1-\beta_e\cosgh)\intd s.
\end{equation}
For a simple model, we set:
\begin{equation}
B(r)\simeq B_\ast\left(\frac{R_\ast}{r}\right)^3, \quad N_e(r)\simeq\eta_e\frac{\Omega B(r)}{2\pi ec}.
\end{equation}
with $B_\ast$ the surface magnetic field.  Thus the optical depth is
given by (e.g. Rafikov \& Goldreich 2005)
\begin{equation}
\tau_- \simeq\frac{2\pi}{3}\eta_e(1-\beta_e\cos\theta_B)
             \frac{r_{e,\rm cyc}}{c/\Omega}
\simeq0.62 \eta B_{\ast12}^{1/3} \gamma_e^{-1/3}\nu_9^{-1/3} 
      P_{\rm 1s}^{-1} (1-\beta_e\cos\theta_B)^{2/3}.\label{eq:tau}
\end{equation}
with $B_{\ast12}=B_\ast/(10^{12}\,{\rm G})$, $\nu_9=\nu/(10^9\,{\rm
  Hz})$. From eq.~(\ref{eq:cycres}) we can find the resonance
radius of the electron
\begin{equation}
r_{e,\rm cyc}/R_\ast 
= 1.8\times10^3B_{\ast12}^{1/3}\nu_9^{-1/3}\gamma_e^{-1/3}\theta_B^{-2/3}.
\label{eq:r_cyc}
\end{equation}

The optical depth of the left-handed circular polarized mode caused by the
scattering of positrons is similarly given by
\begin{equation}
\tau_+ \simeq\frac{2\pi}{3}\eta_p(1-\beta_p\cos\theta_B)\frac{r_{p,\rm cyc}}{c/\Omega}.
\end{equation}
with $\eta_p=N_p/N_{\rm GJ}$ and $r_{p,\rm cyc}$ defined by
eq.~({\ref{eq:r_cyc}) except using $\gamma_p$ instead of $\gamma_e$.

When there is an asymmetry between electrons and positrons (different
density and/or different $\gamma$), the optical depths of the two modes
are different:
\begin{equation}
\Delta \tau = \tau_+-\tau_- = 2\tau \lp\frac{\Delta N}{N} -
\frac{\Delta\gamma}{6\gamma}\rp,
\end{equation}
with $\tau \simeq \tau_+\simeq\tau_-$. 
Now consider a linear-polarized
photon propagating through the cyclotron resonance region. The mode
evolution is non-adiabatic (which is always the case since the
resonance happens after the polarization limiting radius; see sect.~4.2). 
Before the
resonance, the total intensity is
\begin{equation}
I_{\rm i}=I_{\rm i,+}+I_{\rm i, -}, \quad{\rm with}~I_{\rm i,+}=I_{\rm i, -}
\end{equation}
which means that the intensities of the two circular-polarized modes are
the same.  The wave intensity after the cyclotron absorption is
\begin{equation}
I_{\rm f}=I_{\rm f,+}+I_{\rm f,-}=
I_{\rm i,+}e^{-\tau_+}+I_{\rm i,-}e^{-\tau_-}.
\end{equation}
Because of the difference between $\tau_+$ and $\tau_-$, the final
intensities of the two circular-polarized modes are different. Thus
circular polarization can be generated:
\begin{equation}
V_f = I_{\rm f,+} - I_{\rm f,-}=
I_{\rm i,+}e^{-\tau_+}-I_{\rm i,-}e^{-\tau_-}.\end{equation}
\begin{equation}
\frac{V_f}{I_f} = \frac{e^{-\tau_+}-e^{-\tau_-}}{e^{-\tau_+}+e^{-\tau_-}}. \label{eq:ab}
\end{equation}
When $\tau_\pm\ll1$, $V_f/I_f = -\Delta\tau/2 =
\tau\lp\frac{\Delta\gamma}{6\gamma} - \frac{\Delta N}{N}\rp $.

We can also obtain the same result formally by using the Stokes
parameters evolution equation~(\ref{eq:IQUV_evol_singlegg}).
Since electrons and positrons have slightly different $\gamma$,
the cyclotron absorptions caused by electrons and positrons occur at
different radii. We analyse them separately.  Consider the cyclotron
absorption caused by electrons first. Near the resonance,
with
\begin{equation}
x=\frac{r-r_{e,\rm cyc}}{r_{e,\rm cyc}}, \quad |x|\ll 1,
\end{equation}
we have
\begin{equation}
u\gamma_e^{-2}\ggtheta^{-2}\simeq \lp\frac{r}{r_{e,\rm cyc}}\rp^{-6} 
= (1-x)^6\simeq 1-6x,
\end{equation}
where we have assumed $B\propto r^{-3}$. The imaginary part of
$F_{11}$ and $F_{12}$ in eq.~(\ref{eq:IQUV_evol_singlegg}) are
\begin{equation}
F_{11,i}=f_{e, 11,i}={\rm Im}(f_{e,11})
  \simeq {\rm Im}\lp-\frac{v_e\gamma^{-1}}{2i\damp+6x}\rp
  =\frac{v_e\gamma^{-1}}{2\damp}\frac{1}{1+(3x/\damp)^2},
\end{equation}
\begin{equation}
F_{12,i}=f_{e, 12,i}={\rm Im}(f_{e,12})
  \simeq{\rm Im}\lb-f_{e,11}(1+x)^3\rb \simeq -f_{e,11,i}.
\end{equation}
Also $|F_{\eta}|\ll |f_{e,11,i}|$ near the resonance, so we
neglect it.  Thus, the evolution equation for $I$ and $V$ in
eq.~(\ref{eq:IQUV_evol_singlegg}) are simplified to 
\begin{eqnarray}
\intd I/\intd r&=&-k_0f_{e,11,i}I+k_0f_{e,11,i}V,\nonumber\\
\intd V/\intd r&=& k_0f_{e,11,i}I-k_0f_{e,11,i}V. \label{eq:IQUV_evol_cyc}
\end{eqnarray}
Then we have
\begin{eqnarray}
\intd I_+/\intd r&=&0,\nonumber\\
\intd I_-/\intd r&=&-2k_0f_{e,11,i}I_-,
\end{eqnarray}
with $I_+=(I+V)/2$ the intensity of left circular polarized mode and
$I_-=(I-V)/2$ the right one. The solution of these equations is
\begin{eqnarray}
I_{f}&=& I_{i,+}+I_{i,-}e^{-\tau_-}\nonumber\\
V_{f}&=& I_{i,+}+I_{i,-}e^{-\tau_-},
\end{eqnarray}
where $I_{i,+}$, $I_{i,-}$ are the circular-polarized mode intensities
before the resonance, and
\begin{equation}
\tau_-=\int_{\rm across CR} 2k_0f_{e,11,i}\intd r
      =\frac{2\pi}{3}\eta_e(1-\beta_e\cosgh)\frac{r_{e,{\rm cyc}}}{c/\Omega},
\end{equation}
in agreement with eq.~(\ref{eq:tau}).
For the cyclotron absorption by positrons, the analysis is
exactly the same, except $f_{p, 12}=f_{p,11}(1+x)^3 \simeq f_{p,11}$.
The intensities evolution equations are
\begin{eqnarray}
\intd I_+/\intd r&=&-2k_0f_{p,11,i}I_+ ,\nonumber\\
\intd I_-/\intd r&=&0.
\end{eqnarray}
Including both cyclotron absorption by electrons and positrons, the 
intensity and Stokes $V$ parameters after the resonance are
\begin{eqnarray}
I_f&=& I_{i,+}e^{-\tau_+}+I_{i,-}e^{-\tau_-}\nonumber\\
V_f&=& I_{i,+}e^{-\tau_+}-I_{i,-}e^{-\tau_-}
\end{eqnarray}
with
\begin{equation}
\tau_+=\int_{\rm across CR} 2k_0f_{p,11,i}\intd r
      =\frac{2\pi}{3}\eta_p(1-\beta_p\cosgh)\frac{r_{p,{\rm cyc}}}{r_{\rm lc}}.
\end{equation}
Thus, our evolution equations for the mode and Stokes parameters
derived in section 3
automatically include the correct physics of cyclotron absorption by
electrons and positrons.

\subsection{Wave mode coupling}
\label{sec:wmc}

Wave mode coupling happens near the ``polarization limiting radius'',
$r_{\rm pl}$, where the mode evolution changes from adiabatic to
non-adiabatic, i.e., from $\Gamma_{\rm ad}(r<r_{\rm pl}) > 1$ to
$\Gamma_{\rm ad}(r>r_{\rm pl}) < 1$.  Generally, this is caused by the
rotation of the pulsar.  Obviously the concept of wave mode coupling
is relevant for determining the observed polarization only when the
wave mode is linear polarized, i.e. $r_{\rm pl} < r_{\rm cir}$ (see
section 4.3). In the process of wave mode coupling, the circular
polarization will be generated.  For $r<r_{\rm cir}$ (so that
$\theta_m=0$ or $\pi/2$), the mode amplitude evolution equation
(\ref{eq:me}) simplifies to
\begin{equation}
i{\intd\over \intd s}\left(\begin{array}{c}A_+\\A_-\end{array}\right)
=\left(\begin{array}{cc}
-\Delta k/2 & i\phi_B' \\
-i\phi_B' & \Delta k/2
\end{array} \right)
\left(\begin{array}{c}A_+\\A_-\end{array}\right),
\label{eq:me_wmc}
\end{equation}
with $\Delta k=\Delta n \omega/c$.
The adiabatic parameter is defined as
\begin{equation}
\Gamma_{\rm ad}=\left|\frac{\Delta n \omega}{2c\phi_B'}\right|,
\end{equation}
where 
\begin{equation}
\Delta n = \frac{1}{2}\{f_\eta\singh^2-f_{11}[1-\xigh^2]\}.
\end{equation}
When $r< r_{\rm cyc}$, $\Delta n \simeq 1/2f_\eta\singhsq\simeq
-2v\theta_B^{-2}\gamma^{-3}$, so we have
\begin{equation}
\Gamma_{\rm ad}=5.6\times10^9\eta B_{12}\nu_9^{-1}\theta_B^{-2}\gamma^{-3}|F_\phi|^{-1},
\label{eq:Gamma_ad}
\end{equation}
where we have used $\phi_B'=F_\phi/r_{\rm lc}$, $r_{\rm lc}=c/\Omega$, and
\begin{equation}
F_{\phi} = \frac{\sin^2\alpha\cos\zeta-\sin\alpha\cos\alpha\sin\zeta\cos\psi}
{ 1-\left(\cos\alpha\cos\zeta+\sin\zeta\sin\alpha\cos\psi\right)^2}.
\end{equation}
Obviously, $\Gamma_{\rm ad}\gg1$ means adiabatic mode evolution while
$\Gamma_{\rm ad}\ll1$ non-adiabatic. The condition $\Gamma_{\rm
  ad}(r=r_{\rm pl}) = 1$ then gives the polarization limiting radius
\begin{equation}
r_{\rm pl}/R = 1.8\times10^3\eta^{1/3}B_{\ast12}^{1/3}\nu_9^{-1/3}
  \theta_B^{-2/3}\gamma^{-1}|F_\phi|^{-1/3}. \label{eq:rpl}
\end{equation}
Compare $r_{\rm pl}$ with $r_{\rm cyc}$, we have
\begin{equation}
r_{\rm pl}/r_{\rm cyc} = \eta^{1/3}\gamma^{-2/3}|F_\phi|^{-1/3} 
   = 0.215 \eta_2^{1/3}\gamma_2^{-2/3}|F_\phi|^{-1/3}\lo 1. \label{eq:r_pl2r_cyc}
\end{equation}
So in the typcal parameter region ($\eta=100$, $\gamma=100$,
$|F_\phi|=$ a few), wave mode coupling always occurs before cyclotron
absorption.

To understand the wave mode coupling around $r_{\rm pl}$, we write
\begin{equation}
\Gamma_{\rm ad}=x^{-n}, \label{eq:index}
\end{equation}
with $x=r/r_{\rm pl}$. According to eq.~(\ref{eq:Gamma_ad}), the
power-law index $n\sim 3$ (not exactly 3 because $\theta_B$ also
varies as $r$ changes). Then eq.~(\ref{eq:me_wmc}) can be simplified
to
\begin{equation}
i{\intd\over \intd x}\left(\begin{array}{c}A_+\\A_-\end{array}\right)
=|\Lambda|\left(\begin{array}{cc}
x^{-n} & {\rm sign}(\phi_B')i \\
-{\rm sign}(\phi_B')i & -x^{-n}
\end{array} \right)
\left(\begin{array}{c}A_+\\A_-\end{array}\right),
\label{eq:me_wmc_x}
\end{equation}
where
\begin{equation}
\Lambda\equiv r_{\rm pl}\phi_B'= 0.38{\rm
  sign}(\phi_B')\eta^{1/3}B_{\ast12}^{1/3}\nu_9^{-1/3}
  \theta_B^{-2/3}\gamma^{-1}P_1^{-1}|F_\phi|^{2/3}.
\label{eq:Lambda}
\end{equation}
Similar equation is given by van Adelsberg \& Lai (2006), except that
in their paper the dispersion relation of X-ray is dominated by QED
effect so that $\Delta n>0$, while in our case plasma effect dominates
the radio wave propagation with $\Delta n<0$.
Figure~\ref{fig:wmc_single} shows two examples of mode evolution with
$\Lambda=0.1$ and $\Lambda=1.0$, both for $n=3$.  The photon is
$100\%$ linear polarization before the wave mode coupling (Here we set
it to be O-mode initially). After the wave mode coupling ($x\gg1$),
the polarization states are frozen. In this process, circular
polarization is produced. It is obvious that the larger $\Lambda$ is,
the more circular polarization will be generated.
Figure~\ref{fig:GGn} shows how the value of $\Lambda$ affects the
final circular polarization $|V|/I$ when the power index $n=3$.  For
$n=3$ and $\Lambda<0.1$, the final circular polarization is given by
the expression
\begin{equation}
V/I=2.2{\rm sign}(\phi_B')|\Lambda|^{3/2} \label{eq:GG_n3}.
\label{eq:VI}\end{equation}
For $\Lambda\go1$, the circular polarization $|V|/I$ is close to 1.
Equation~(\ref{eq:GG_n3}) also shows the relationship between the sign
of the circular polarization and $\phi_B'$.  An increasing $\phi_B$
(or $\phi_B'>0$) corresponds to positive circular polarization while
decreasing $\phi_B$ to negative one.

\begin{figure}
\centerline{\psfig{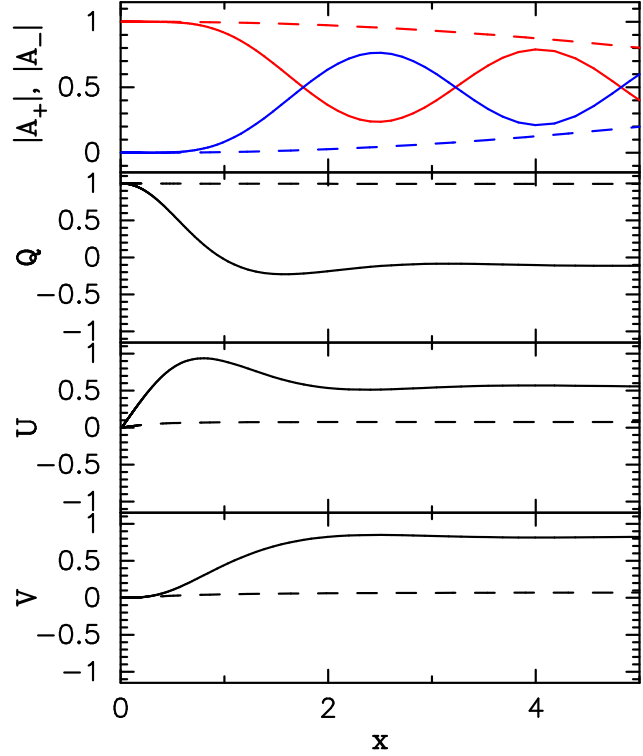}}
\caption{Evolution of the radiation mode amplitudes (top panel) and
  Stokes parameters (bottom three panels) with power-law index $n=3$
  [see eq.~(\ref{eq:index})].  The solid lines are for $\Lambda=1.0$
  and the dashed lines for $\Lambda=0.1$. The polarization limiting
  radius is at $x=1$. The initial values (at a small $x=x_{\rm i}$)
  are $A_+=1$, $A_-=0$, $Q=I=1$, $U=0$ and $V=0$. When $x\lo 0.5$, the
  modes evolve adiabatically.  At $r\sim r_{\rm pl}$ (or $x=1$) the
  modes begin to couple, generating circular polarization. At $x\gg
  1$, the Stokes parameters are ``frozen''.}\label{fig:wmc_single}
\end{figure}

\begin{figure}
\centerline{\psfig{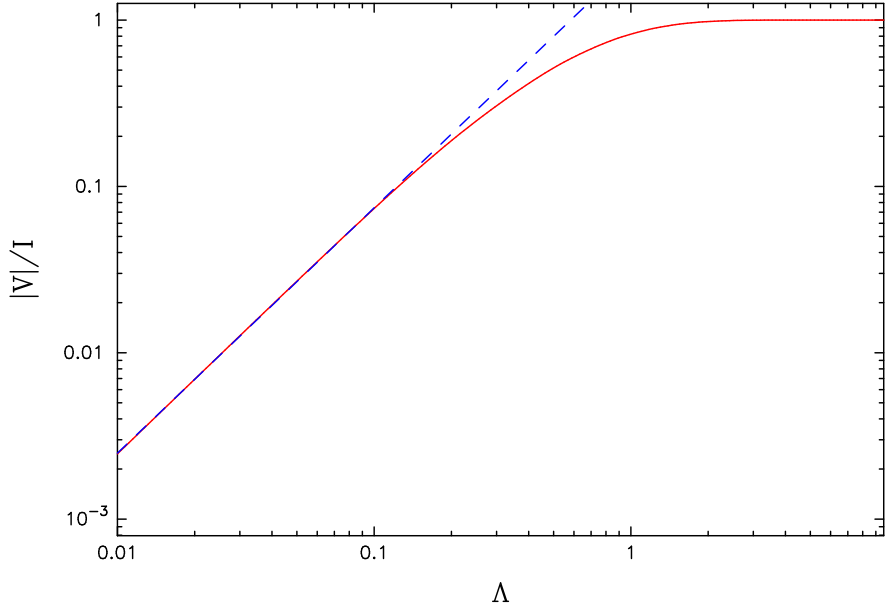}}
\caption{The final circular polarization fraction $|V|/I$ after wave
  mode coupling as a function of $\Lambda$ with the power-law index
  $n=3$. The linear polarization fraction before wave mode coupling is
  assumed to be $100\%$. The dash line depicts the fitting formula
  [eq.~(\ref{eq:GG_n3})] for $\Lambda<0.1$.  }\label{fig:GGn}
\end{figure}

\subsection{Circularization}

Circularization happens when $|\polarb|\sim 1$, and we can define the
radius of circularization $r_{\rm cir}$ by
$|\polarb(r=r_{\rm cir})| = 1$. For $r\gg r_{\rm cir}$, the normal modes
become circular-polarized.

According to eq.~(\ref{eq:polarb_beforecyc}), if $r_{\rm cir}\ll
r_{\rm cyc}$ (or $\lambda\gg 1$, before cyclotron resonance) and
$\theta_B\gamma\gg 1$, the polarization parameter
\begin{equation}
|\polarb| \simeq \frac{\lambda}
{\theta_B^2\gamma^2\Delta N/N-\Delta\gamma/\gamma} \gg 1, \label{eq:beta_pol_1}
\end{equation}
which means the two wave modes are always linear polarized. However,
if $r_{\rm cir}\ll r_{\rm cyc}$ and $\theta_B\gamma\ll 1$,
\begin{equation}
|\polarb| \simeq \frac{\lambda\theta_B^2\gamma^2}
{\Delta N/N-\Delta\gamma/\gamma},\label{eq:beta_pol_2}
\end{equation}
so circularization could happen when
\begin{equation}
\theta_B\gamma = \sqrt{(\Delta N/ N - \Delta\gamma/\gamma)/\lambda}.
\label{eq:theta_cir}
\end{equation}
Which implies a very small $\theta_B$. This condition could be
satisfied when the photon ray is nearly aligned with the magenetic
field, or when the photon is generated inside the $1/\gamma$ cone of
the radiation beam.

If the circularization happens after the cyclotron resonance ($r_{\rm
  cir}\gg r_{\rm cyc}$ or $\lambda \ll 1$), according to
eq.~(\ref{eq:polarb_aftercyc}), the radius of circularization is given
by
\begin{equation}
r_{\rm cir}/R_\ast= 2.2\times10^3B_{\ast12}^{1/3}\nu_9^{-1/3}\theta_B^{-4/3}
\gamma^{-1/3}(\Delta N/N-\Delta\gamma/\gamma)^{-1/3}. \label{eq:r_cir}
\end{equation}
Here we used $B \simeq B_\ast(r/R_\ast)^{-3}$ and assumed
$\theta_B\gamma\gg 1$. The ratio of $r_{\rm cir}$ and the cyclotron
resonance radius [see eq.~(\ref{eq:r_cyc})] is
\begin{equation}
r_{\rm cir}/r_{\rm cyc}= 1.2~\theta_B^{-2/3}(\Delta N/N-\Delta\gamma/\gamma)^{-1/3}\gg 1.\label{eq:r_cir2r_cyc}
\end{equation}
Obviously, in the parameter regions we interested in, the
circularization radius is typically larger than the cyclotron
resonance radius and the polarization limiting radius. Thus, this
effect does not change the photon polarization state at all.

\subsection{Quasi-Tangential Propagation Effect}

In their study of the X-ray polarization signals from magnetized
neutron stars, Wang \& Lai (2009) found that as the X-ray photon
travels through the magnetosphere, it may cross the region where its
wave vector is aligned or nearly aligned with the magnetic field
(i.e., $\theta_B$ is zero or small). In such a {\it Quasi-Tangential}
region (QT region), the azimuthal angle of magnetic field $\phi_B$
changes quickly, the two photon modes ($\parallel$ and $\perp$ modes)
become (nearly) identical, and mode coupling may occur, thereby
affecting the polarization alignment. This Quasi-tangential Effect
generally happens at a few $R_\ast$ for surface X-ray emission. The
physical mechanism is similar to the wave mode coupling effect
discussed in section 4.2 [see the mode evolution equation~(2.11) in
  Wang \& Lai (2009)], except that the magnetic field plays an
important role.

In the radio case, we assume the photon is emitted in the tangential
direction of the magnetic field line at the emission point ($\sim
50R_\ast$). If the NS is non-rotating, then the $\veck-\vecB$ angle
$\theta_B$ ($=0$ at the emission point) will increase monotonically
and no QT effect will occur. However, when we consider the rotation of
the NS, for some special photons (for example, those with small impact
angle $\chi$ and special $\Psi_{\rm i}$) $\theta_B$ could attain its
minimum value at a large radius. As an example, the two bottom panels
of Figure~\ref{fig:single_qt_1} and \ref{fig:single_qt_2} (to be
discussed in detail in section 5) shows the evolution of $\theta_B$,
$\phi_B$ along the ray for $\chi=0.5^o$. We see that $\theta_B$
reaches its minimum value at about $s=700R_\ast$ away from the
emission point. The azimuthal angle $\phi_B$ changes very quickly at
this radius. The two linear modes strongly couple with each other. The
final polarization state after crossing this QT region is complicated:
$\phi_{\rm PA}$ can be modified significantly and different sign of
circular polarization can be generated for different geometry, which
is different from wave mode coupling effect discussed in section
4.2. In general, the QT effect strongly influence the polarization
phase profiles when impact angle is very small (see section 5.3).  In
our case, the QT effect is always coupled with wave mode coupling
effect (occurring at almost same place), and the numerical ray
integration is necessary to account for these effects accurately (see
section 5).

\section {Numerical Results}

In section 4 we have discussed various key physical effects related to
wave propagation through the magnetosphere. However, in many cases
these different effects are coupled and not easy to separate. Thus, to
produced the observed polarization profiles, it is necessary to use
the numerical ray integrations to calculate the final wave polarization
states.

\subsection{Single Ray evolution}
\label{sec:single}

It is generally accepted that pulsar radio emission is emitted from
the open field line region at a few to tens of NS radii (e.g. Cordes
1978; Blaskiewicz et al. 1991; Kramer et al.~1997; Kijak \& Gil
2003). In this paper, we choose the emission height $r_{\rm
  em}=50R_\ast$ and assume that at the emission point, the photon is
polarized in the $\veck$-$\vecB$ plane (or the O-mode, as in the case
of curvature radiation), and propagates along the tangential direction
of the local magnetic field line (here we do not consider the emission
cone of angle $1/\gamma$). For a given emission height $r_{\rm em}$,
the pulsar rotation phase $\Psi_i$, the direction of line of sight
$\zeta$ (which is the $\veck$-$\vecW$ angle), the surface magnetic
field $B_\ast$, and the plasma properties (plasma density parameter
$\eta$, Lorentz factor of the streaming plasma $\gamma$), we can
calculate the dielectric tensor at each point along the photon ray,
and integrate the wave evolution equation~(\ref{eq:evol}) from the
emission point to a large radius (generally we choose $r_{\rm lc}/2$),
beyond the polarization limiting radius $r_{\rm pl}$ and cyclotron
resonance radius $r_{\rm cyc}$, to determine the final polarization
state of the photon.

\subsubsection{Symmetric pair plasma}

We first consider the case of symmetric pair plasmas, i.e., the
electrons and positrons have the same Lorentz factors
($\gamma_p=\gamma_e$, or $\Delta\gamma/\gamma =0$) and densities
($\Delta N/N =0$). In this case, the eigenmodes are always linear
polarized (mixing angle $\theta_m=0^o$ or $90^o$).
Figure~\ref{fig:single_1} shows an example of the photon polarization
evolution along its trajectory.  We can clearly find the wave mode
coupling effect (at $r_{\rm pl}\sim$ 800\,$R_\ast$) and cyclotron
absorption effect (at $r_{\rm cyc}$ more than 1000$R_\ast$). The final
polarization position angle $\phi_{\rm PA}$ is determined by
$\phi_B(r_{\rm pl})$ [see eq.~(\ref{eq:phi_rpl})].  It is obvious that
near the polarization limiting radius, $\Gamma_{\rm ad}\propto
(r/r_{\rm pl})^{-n}$ and $n\sim3$, so that as discussed in
section~\ref{sec:wmc}, the final circular polarization is determined
by the value of $\Lambda$ [see eq.~(\ref{eq:GG_n3})]. Since we are
dealing with a symmetric pair plasma here, cyclotron absorptions do
not change the polarization state (but decrease the total intensity).

Figure~\ref{fig:single_all} give some other examples of the evolution
of Stokes parameters with different plasma density $\eta$ and Lorentz
factor $\gamma$. Different $\eta$ and $\gamma$ correspond to different
$r_{\rm pl}$ [according to eq.~(\ref{eq:rpl}), lower $\eta$ and higher
  $\gamma$ corresponds to a smaller $r_{\rm pl}$], so that the final
$\phi_{\rm PA}$ is different too. In all the above cases, the final
polarization state changes significantly compared to the original
state, not only the linear position angle but also the circular
polarization.

At the special parameter region of the initial rotation phase
$\Psi_{\rm i}$, the QT effect (see section 4.4) can strongly affect
the final polarization state.  Figure~\ref{fig:single_qt_1} shows the
photon evolution for $\Psi_{\rm i}=-9^o$ (the other parameters are the
sames as in Fig.~\ref{fig:single_1}, e.g. the impact angle
$\chi=2^o$).  Note that in contrast to Fig.~\ref{fig:single_1}, here
the $\veck-\vecB$ angle $\theta_B$ does not vary monotonically along
the ray. There exists a QT region around $s\sim700R_\ast$, where
$\theta_B$ is minimum and $\phi_B$ is changing very quickly.  As
discussed in section~4.4, the final $\phi_{\rm PA}$ and circular
polarization are different from the prediction of pure wave mode
coupling effect (which is the case in Fig.~\ref{fig:single_1} where QT
effect does not occur).  For the photon evolution with a smaller
photon impact angle $\chi=0.5^o$ (but the initial rotation phase and
other parameters are the same as in Fig.~\ref{fig:single_qt_1}), the
QT effect is stronger, as shown in Figure~\ref{fig:single_qt_2}. Note
that even the sign of the final circular polarization in this figure
is positive, as a result of the strong QT effect.

\begin{figure}
\centerline{\psfig{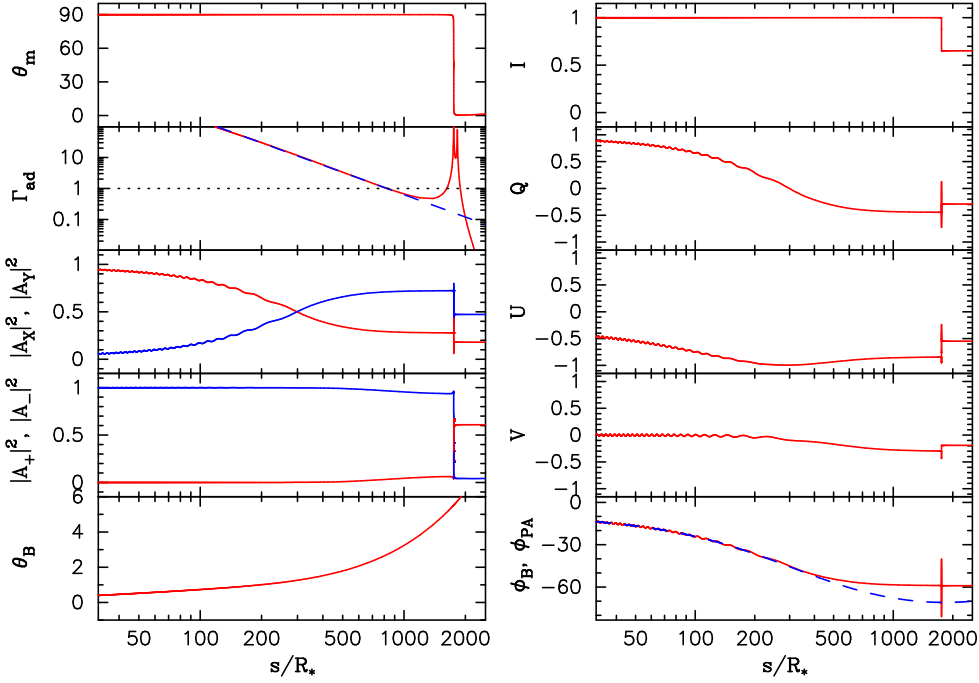}}
\caption{ An typical single photon evolution across the magnetosphere.
  The horizontal axis $s/R_*$ is the photon distance away from the
  emission point. On the left panels, $\theta_{\rm m}$ is the mode
  mixing angle defined by eq.~(\ref{eq:theta_m}), $\Gamma_{\rm ad}$ is
  the adiabatic parameter defined by eq.~(\ref{eq:Gamma_ad}) (the
  dashed line is the power-law fit of $\Gamma_{\rm ad}$ around $r_{\rm
    pl}$, which is $\Gamma_{\rm ad}\propto s^{-2.39}$), $A_X$ and
  $A_Y$ are the wave amplitudes in the fixed $XYZ$ frame.  $A_+$ and
  $A_-$ are the mode amplitudes, and $\theta_B$ is the angle between
  $\veck$ and $\vecB$. On the right panels, $I$, $Q$, $U$, $V$ are the
  Stokes parameters, $\PA=0.5\tan^{-1}(U/Q)$ is the linear
  polarization position angle (solid line) and $\phi_B$ is the
  azimuthal angle of the $\vecB$ field (dashed line). The initial
  polarization is assumed to be in the ordinary mode, with $A_+=1$,
  $A_-=0$. For this example, the parameters are: surface magnetic
  field $B_\ast=10^{12}$\,G, NS spin period $P=1$\,s, wave frequency
  $\nu=1$\,GHz, plasma density parameter $\eta=N/N_{\rm GJ}=400$
  ($N=N_e+N_p$ and $N_e=N_p$), Lorentz factor $\gamma=100$ (with
  $\Delta\gamma/\gamma\simeq 0$), inclination angle $\alpha=30^o$,
  impact angle $\chi=2^o$, initial rotation phase $\Psi_{\rm i}=0^o$,
  and emission height $r_{\rm em}=50R_\ast$. \label{fig:single_1}}
\end{figure}

\begin{figure}
\centerline{\psfig{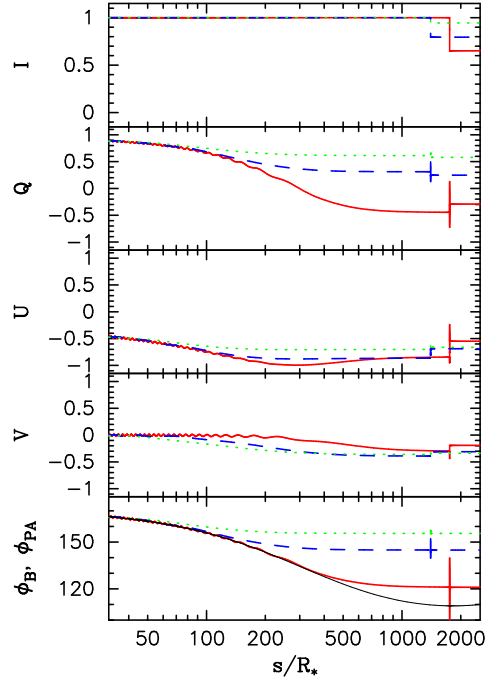}}
\caption{ Same as the right panels of Fig.~\ref{fig:single_1}, except
  for different plasma density $\eta$ and Lorentz factor of the streaming
  motion $\gamma$: the solid lines are for $\eta=400$, $\gamma=100$
  (same as Fig.~\ref{fig:single_1}), the dashed lines for $\eta=400$,
  $\gamma=300$ and the dotted lines for $\eta=100$, $\gamma=100$. The
  thin line in the bottom panel is for $\phi_B$. It is obvious that for
  lower density and/or higher $\gamma$, the wave mode coupling occurs
  at smaller $r_{\rm pl}$. \label{fig:single_all} }
\end{figure}

\begin{figure}
\centerline{\psfig{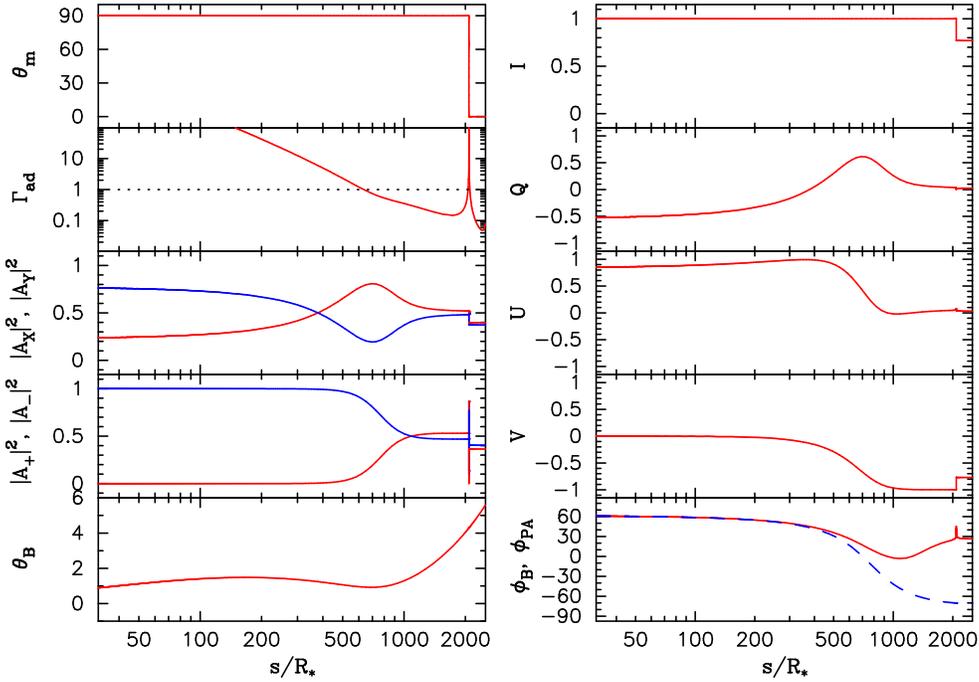}}
\caption{ Same as Fig.~\ref{fig:single_1}, except for a different
  initial rotation phase $\Psi_{\rm i}=-9^o$.  The impact angle is the
  same as in Fig.~\ref{fig:single_1}, $\chi=2^o$.  Note that in
  contrast to Fig.~\ref{fig:single_1}, here the $\veck-\vecB$ angle
  $\theta_B$ does not vary monotonically along the ray.  There exists
  a quasi-tangential (QT) region around $s\sim700R_\ast$, where
  $\theta_B$ is minimum and $\phi_B$ changes very quickly. Strong
  circular polarization is generated here and the final PA angle
  $\phi_{\rm PA}$ can not be predicted by simply using
  eq.~(\ref{eq:PA}).
 \label{fig:single_qt_1}}
\end{figure}

\begin{figure}
\centerline{\psfig{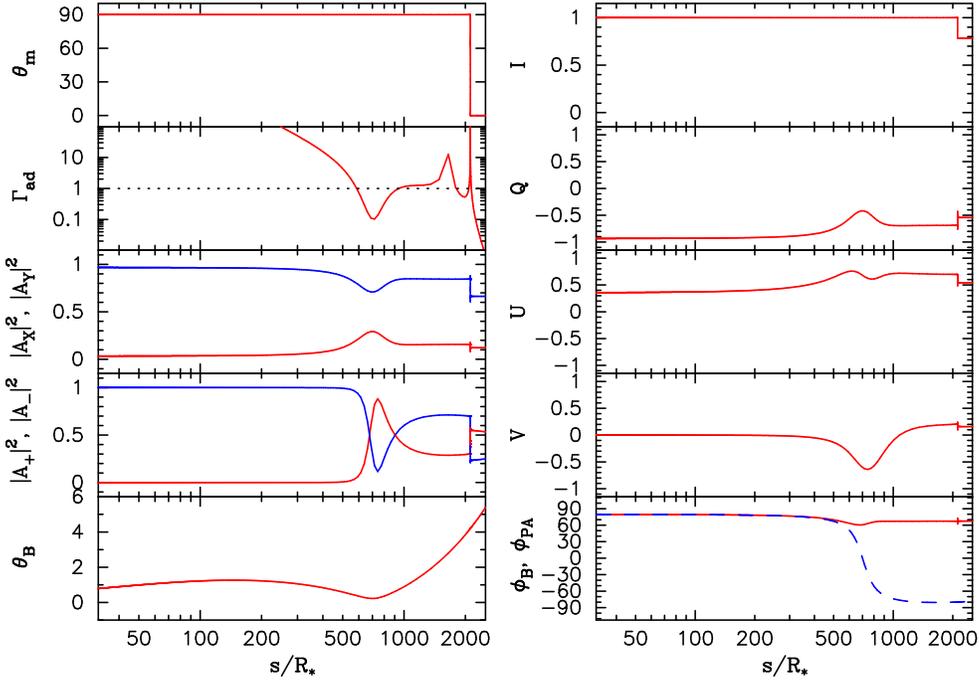}}
\caption{ Same as Fig.~\ref{fig:single_qt_1}, except for the impact
  angle $\chi=0.5^o$. The very smaller impact angle makes $\phi_B$
  change more quickly around $s\sim700R_\ast$, so that the QT effect
  is stronger than the case shown in Fig.~\ref{fig:single_qt_1}. Note
  that the sign of final circular polarization is different from the
  cases in Fig.~\ref{fig:single_1} and \ref{fig:single_qt_1}.
 \label{fig:single_qt_2}}
\end{figure}

\subsubsection{Asymmetric pair plasma}

If the electrons and positrons of the magnetospheric plasma have
different velocities and/or densities, the wave eigenmodes cannot
always be linearly polarized.  As discussed in section 4.3, before
cyclotron resonance the natural modes are linearly polarized [see
  eq.~(\ref{eq:beta_pol_1})] for $\theta_B\gamma\gg1$. After the
cyclotron resonance, the natural modes become elliptical polarized. In
section 4.3, we have defined a circularization radius $r_{\rm cir}$
where the polarization parameter $|\beta_{\rm pol}|=1$ [see
  eqs.~(\ref{eq:r_cir}) and (\ref{eq:r_cir2r_cyc})]. For $r\gg r_{\rm
  cir}$, the natural modes becomes circular polarized.

According to eq.~(\ref{eq:r_pl2r_cyc}), for typical plasma parameters
of interest in this paper, $\eta\go 100$ and $\gamma\go 100$, wave
mode coupling always occurs before the cyclotron resonance ($r_{\rm
  pl}<r_{\rm cyc}$). Thus, circularization always happens after wave
mode coupling, at which point the wave polarization state is already
frozen.  Therefore the change of natural mode does not affect the
observed polarization state. Figure~\ref{fig:single_asym} shows the
photon evolution in an asymmetric pair plasma. Note that the mode
mixing angle $\theta_m$ changes from $0^o/90^o$ to $45^o$ after the
cyclotron resonance, but the polarization state does not change since
$r_{\rm cir}>r_{\rm pl}$.

von Hoensbroech et al. (1998) studied wave modes in a pure electron
plasma. They assumed that the background plasma has a much lower
Lorentz factor (e.g. $\gamma_{\rm bg}=1.7$) than the Lorentz factor of
the radiating beam. In this case $r_{\rm cir}$ may be close to $r_{\rm
  pl}$ and circular polarization may be generated around $r_{\rm
  cir}$.  Note that they did not calculate $r_{\rm pl}$ but simply
assumed that the final photon polarization is determined by the normal
mode at some fixed $r_{\rm pl}$.

\begin{figure}
\centerline{\psfig{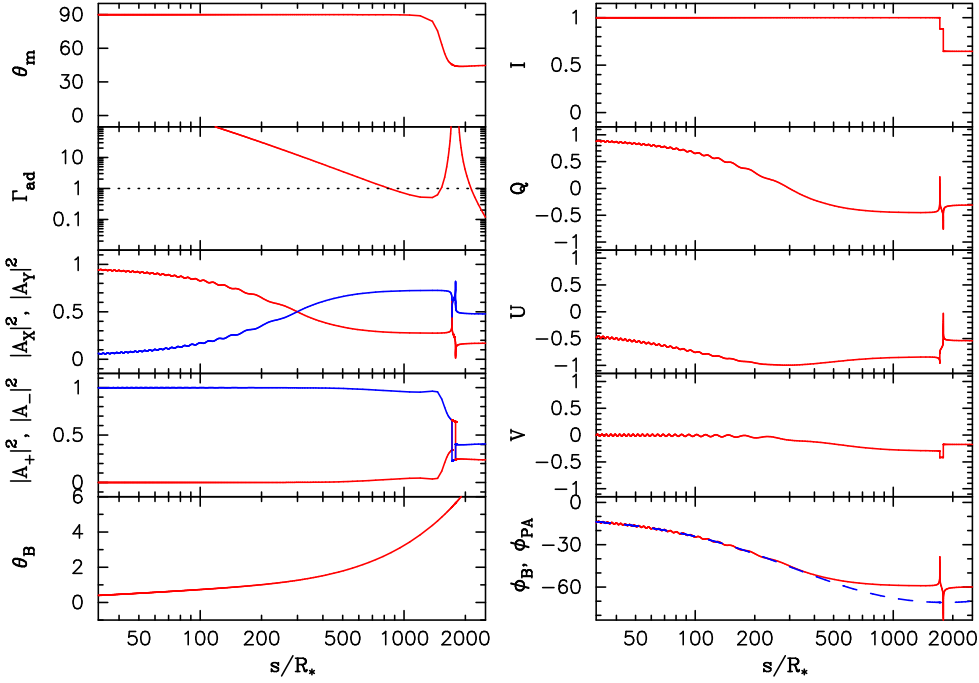}}
\caption{ Same as Fig.~\ref{fig:single_1}, except for an asymmetric
  pair plasma with $\Delta \gamma/\gamma=0.2$, $\Delta N/N=0$.  Since
  the Lorentz factors of electrons and positrons are different, their
  cyclotron resonance occur at different radii.  Note that after the
  cyclotron resonance, the natural modes become circular polarized
  ($\theta_m=45^o$).  \label{fig:single_asym}}
\end{figure}


\subsection{Polarization Profiles of Pulsar Emission Beam}

Having understood the main features of polarization evolution along a
single ray, we now proceed to calculate the polarization profiles of
pulsar emission beam. To do this, one needs to know the emission
height as a function of pulsar rotation phase. For simplicity, in this
paper, we assume that all emissions are from the same height, at
$r_{\rm em}=50R_\ast$, and defer the results for emissions from a
range of heights to a future paper. For a given emission height
$r_{\rm em}$, the pulsar rotation phase $\Psi_i$, the inclination
angle $\alpha$ and the direction of line of sight $\zeta$ (which is
the $\veck$-$\vecW$ angle), we can find the position of the emission
point $\vecr_{\rm i}$ where the tangential magnetic field line
direction is along the line of sight. This emission point, $\vecr_{\rm
  i}=(r_{\rm em},~\theta_\ri,~\phi_\ri)$, is given by (in the fixed
$XYZ$ frame)
\begin{equation}
\theta_\ri=\frac{\theta_\mui}{2}-\frac{1}{2}\sin^{-1}\left(\frac{1}{3}\sin\theta_\mui\right),\quad~~
\phi_\ri=\phi_\mui, \label{eq:theta_ri}
\end{equation}
where ($\theta_\mui$, $\phi_\mui$) is the initial direction of the
dipole magnetic momentum $\vecmu_i$ and can be found in
eq.~(\ref{eq:theta_mu}) (with $\Psi$ given by $\Psi_{\rm i}$).  We
consider emissions only from the open field line region, i.e., the
angle between $\vecr_{\rm i}$ and $\vecmu_{\rm i}$ should be less than
$\sqrt{r_{\rm em}/r_{\rm lc}}$.

For given $r_{\rm em}$, $\alpha$, $\zeta$, $\Psi_i$ and initial
polarization state (ordinary mode), we determine $\vecr_{\rm i}$ and
calculate the final observed Stokes parameters by integrating along
the ray. When the phase $\Psi_i$ varies due to NS rotation, we can
observe photons from different emission points and the final observed
Stokes parameters will change with the rotation phase --- this is the
pulsar polarization profile. If we neglect the propagation effect, the
observed position angle $\phi_{\rm PA}$ can be described by the
Rotating-Vector-Mode (RVM) (see Radhakrishnan \& Cooke 1969) as
\begin{equation}
  \PA=\phi_\mui=\phi_\mu(\Psi_i)=\tan^{-1}\frac{-\sin\alpha\sin\Psi_i}
      {\sin\zeta\cos\alpha-\cos\zeta\sin\alpha\cos\Psi_i}. \label{eq:PA}
\end{equation}
The basic assumption of the RVM is that the radiation is emitted with
polarization in the plane of the field line curvature (i.e. the
$\veck$-$\vecB$ plane) and this polarization direction is unchanged
during the propagation. However, as seen in section~\ref{sec:single},
the final polarization state can be modified compared to the initial
one because of the propagation effect in the magnetosphere, so that
the final PA profile can deviate significantly from the RVM model.

Figure \ref{fig:1D_diffeta} shows a typical example of the phase
evolution of the intensity and polarization, taking into account of
all the propagation effects. The total intensity is only affected by
cyclotron absorption, and a higher plasma density leads to stronger
absorption. We see that the relative intensity $I/I_0$ varies with the
rotation phase $\Psi_{\rm i}$, simply because the wave passes through
different paths in the magnetosphere for different $\Psi_{\rm i}$.
For illustrative purpose, we consider the initial intensity profile
$I_0$, given by a Gaussian centered at $\Psi_{\rm i}=0$:
\begin{equation}
I_0(\Psi_{\rm  i})=\exp(-4\sqrt{\ln2}\Psi_{\rm i}^2/(\Psi_{\rm max}^2)).
\end{equation}
Here $\Psi_{\rm max}$ is the initial phase of the photon
from the edge of the open field region and is given by
\begin{equation}
\cos\Psi_{\rm max}=\frac{\cos\theta_{\rm open}-\cos\zeta\cos\alpha}{\sin\zeta\sin\alpha}
\end{equation}
where $\theta_{\rm open}\simeq\sqrt{r_{\rm em}/r_{\rm lc}}$ is the
half-cone angle of the open field region at emission height $r_{\rm
  em}$ (here we simply assume the open field region is always the same
as the $\vecmu\parallel\vecW$ case).  Since $I/I_0$ depends
asymmetrically on $\Psi_i$, the observed intensity $I$ is no longer a
Gaussion.  Non-gaussion profiles have been observed in many pulsars,
and the phase-dependent cyclotron absorption illsutrated here is a
possible explanation.

The final polarization profiles are also strongly affected by the
propagation effects. When the plasma density is not so high, and/or
the impact angle $\chi$ is not so small [compared to the half cone
  angle of the emission beam from the open field region $\theta_{\rm
    beam}$; e.g., in Fig.~\ref{fig:1D_diffeta}(a), $\chi=5^o$ while
  $\theta_{\rm beam}\simeq1.5\theta_{\rm open} = 8.8^o$], the wave
mode coupling effect is not strong and the final circular polarization
is not very high. In this case, the final linear polarization position
angle is determined by the azimuthal angle of $\vecB$ field at the
polarization limiting radius $\phi_B(r_{\rm pl})$:
\begin{equation}
\phi_{\rm PA}\simeq\phi_B(r_{\rm pl})\simeq\pi+\phi_\mu(r_{\rm pl})
=\pi+\phi_\mu(\Psi_i+r_{\rm pl}/r_{\rm lc}). \label{eq:phi_rpl}
\end{equation}
Here we have used the approximation of $\vecB(\vecr)=-\vecmu/r^3$
since $r_{\rm pl}\gg R_\ast$.  In general, the polarization limiting
radius, $r_{\rm pl}$, does not vary significantly with different
rotation phase $\Psi_{\rm i}$. Thus the final PA profile just shifts
by the amount
\begin{equation}
r_{\rm pl}/r_{\rm lc}\simeq 0.08 
\lp\frac{\eta}{10^3}\rp^{1/3}
\lp\frac{B}{10^{12}\,{\rm G}}\rp^{1/3}
\lp\frac{\nu}{1\,{\rm GHz}}\rp^{-1/3}
\lp\frac{\gamma}{10^2}\rp^{-1}
\lp\frac{\theta_B}{0.1}\rp^{-2/3}
\lp\frac{P}{1\,{\rm s}}\rp^{-1}
\lp\frac{F_{\phi}}{10}\rp^{-1/3} \label{eq:PAshift}
\end{equation}
compared to the RVM model.  The final circular polarization is always
single sign in this case. We can easily find the relationship between
$\phi_{\rm PA}$ and the sign of circular polarization. According to
eq.~(\ref{eq:phi_rpl}), the monotonicity of the final PA angle
$\phi_{\rm PA}(\Psi_{\rm i})$ is given by
\begin{equation}
\frac{\intd \phi_{\rm PA}}{\intd \Psi_{\rm i}}
   =\frac{\intd \phi_B(r_{\rm pl})}{\intd \Psi}
    \frac{\intd \Psi}{\intd \Psi_{\rm i}}
    =\frac{\intd \phi_B(r_{\rm pl})}{\intd \Psi}.
\end{equation}
Here $\Psi=\Psi_{\rm i}+s/r_{\rm lc}$ [see eq.~(\ref{eq:Psi})], so that
$\intd \Psi/\intd \Psi_{\rm i}=1$.  The sign of the final circular
polarization (generated by wave mode coupling effect) is determined by
$\intd \phi_B(r_{\rm pl})/\intd s$ [$\phi_B'$ in eq.~(\ref{eq:GG_n3})] and:
\begin{equation}
\frac{\intd \phi_B(r_{\rm pl})}{\intd s}
   =\frac{\intd \phi_B(r_{\rm pl})}{\intd \Psi}
    \frac{\intd \Psi}{\intd s}
   =\frac{\intd \phi_B(r_{\rm pl})}{\intd \Psi}\frac{1}{r_{\rm lc}}. \label{eq:PAsign}
\end{equation}
So that $\intd \phi_{\rm PA}/\intd\Psi_{\rm i}$ always has the same
sign as $\intd \phi_B(r_{\rm pl})/\intd s$. According to eqs.~
(\ref{eq:GG_n3}) and (\ref{eq:PAsign}), we can find that because of
the wave mode coupling effect, a monotonically increasing $\phi_{\rm
  PA}$ leads to a positive $V$ while a monotonically decreasing
$\phi_{\rm PA}$ gives a negative $V$.  This relationship has been
observed in some conal-douple type pulsars; see section 5.3.

\begin{figure}
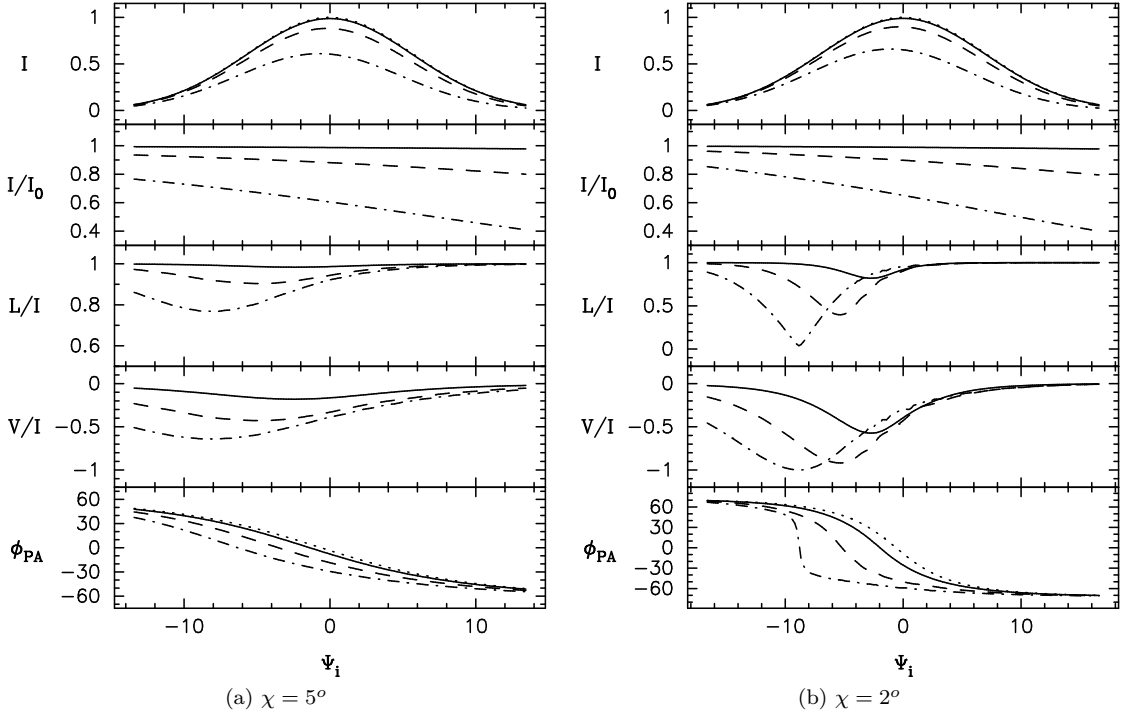

\begin{tabular}{cc}
\psfig{figure=f9a.ps,angle=-90,height=9cm} &
\psfig{figure=f9b.ps,angle=-90,height=9cm} \\
(a) $\chi=5^o$ & (b) $\chi=2^o$\\
\end{tabular}
\caption{The intensity and polarization profiles computed by ray
  integrations.  We use two different impact angles: (a) $\chi=5^o$ on
  the left panels, and (b) $\chi=2^o$ on the right panels.  The solid
  lines are for plasma density parameter $\eta=10$, the dashed lines
  for $\eta=100$ and the dot-dashed line for $\eta=400$.  The top
  panels show the total intensity profiles, where we have adopted (for
  illustrative purpose) a Gaussian initial intensity profile
  $I_0(\Psi_{\rm i})=\exp(-4\sqrt{\ln2}\Psi_{\rm i}^2/\Psi_{\rm
    max}^2))$ (this initial profile is shown shown as dotted lines,
  almost coincident with the solid lines).  The second panels form the
  top show the modification factor $I/I_0$ due to propagation
  effects. The bottom three panels show the linear polarization
  fraction $L/I$, circular polarization $V/I$ and the position angle
  of the LP, $\PA$. In the bottom panels, the dotted lines (almost
  coincident with the solid line on the left panel) show the
  prediction from the RVM model: $\PA =\phi_\mui\simeq\phi_{Bi}$.  In
  these calculations, the initial polarization states are all ordinary
  mode, and the other parameters are: surface magnetic field
  $B_\ast=10^{12}$\,G, NS spin period $P=1$\,s, wave frequency
  $\nu=1$\,GHz, Lorentz factor $\gamma=100$ (with
  $\Delta\gamma/\gamma\simeq 0$), inclination angle $\alpha=30^o$, and
  emission height $r_{\rm em}=50R_\ast$.}\label{fig:1D_diffeta}
\end{figure}

The polarization profiles can also be quite different from the RVM
prediction, especially in the case of low impact angle and/or high
plasma density.  Figure~\ref{fig:1D_diffeta}(b) give some examples for
the impact angle $\chi=2^o$. For the low density case of $\eta=10$,
the final PA profile can still be approximated by a simple shift from
the RVM model. However, for higher density ($\eta=200$, 400), the
final PA profile is not just a simple shift compared with the RVM
model. For example, the PA profile of $\eta=400$ case has a $90^o$
jump within $1^o$ around $\Psi_i\simeq -9^o$, where the linear
polarization $L/I$ is close to 0 while the circular polarization
$|V|/I$ reaches almost $100\%$.  In this region the QT effect
(discussed in section 4.4) plays an important role in determining the
final polarization state (see Fig.~\ref{fig:single_qt_1}).

\subsection{Two-Dimensional Polarization Maps of Pulsar Emission Beam}

For a given pulsar, observation at different line of sight (i.e.,
different $\zeta$ or impact angle $\chi$) would obviously result in
different intensity and polarization profiles.  Figure~\ref{fig:2D_1}
gives an example of the two-dimensional polarization map of the
observed Stokes parameters, produced by varying $\chi$ and $\Psi_i$,
while keeping all other parameters fixed.  As discussed before, the
final total intensity $I$ is only affected by cyclotron absorption,
while the linear and circular polarizations are modified by wave mode
coupling effect and QT effect, and can deviate significantly from the
prediction of RVM model. Figure~\ref{fig:2D_4t} shows four profiles with
four different impact angle $\chi$, corresponding to four sections of
Fig.~\ref{fig:2D_1}. These four sections represent three typical final
polarization states produced by the propagation effects:

(i) For a relatively large impact angle $|\chi|$, the final PA profile
can be obtained by a small shift from the RVM profile, of the amount
$r_{\rm pl}/r_{\rm lc}$ [see eqs.~(\ref{eq:phi_rpl}) and
  (\ref{eq:PAshift})].  Figure~\ref{fig:1D_diffeta} and the
$\chi=-5^o$ column of Fig.~\ref{fig:2D_4t} depict some examples. The
final circular polarization is always of single sign: a monotonically
increasing $\phi_{\rm PA}$ leads to a positive $V$ while a
monotonically decreasing $\phi_{\rm PA}$ gives a negative $V$ [see
  eqs.~(\ref{eq:GG_n3}) and (\ref{eq:PAsign})].  This behavior is
consistent with observations of the double cone emission of some
pulsars (``conal-double type pulsars''), where a correlation between
the sense of CP and the sense of PA variation was found (see Han et
al.~1998).

(ii) For relatively small impact angle, the final PA profile is very
different from the RVM prediction -- the middle two columns
($\chi=-1.9^0$ and $\chi=-1^0$) of Fig.~\ref{fig:2D_4t}) give some
examples.  It is clear that there always exists a special line of
sight ($\chi=\chi_{\rm jump}$), for which the PA profile has a $90^o$
jump (where $Q=0$ while $U$ changes signs [see the $\chi=-1.9^0$
  column of Fig.~\ref{fig:2D_4t}]. The large jumps in $V/I$ and
$\phi_{\rm PA}$ are caused by the QT effect. For $|\chi|<\chi_{\rm
  jump}$, the PA is not necessarily an monotonic function of $\Psi_i$.
Nevertheless, the final CP retains a single sign, which is the same as
the case with large $|\chi|$.

(iii) For very small impact angle ($|\chi|\ll \chi_{\rm jump}$), the
QT effect is much stronger, so that the final PA profile is very
different from the prediction of RVM model and even the CP does not
stay at a single sign (see the right-most column of
Fig.~\ref{fig:2D_4t}).

The above three types of polarization behaviours always exist for
different pulsar and plasma parameters (e.g., $B_\ast$, $P_0$, $\nu$,
$\eta$, $\gamma$, $\alpha$ and $r_{\rm em}$). Different parameters
just modify the position of $\chi_{\rm jump}$ and the initial rotation
phase where the $90^o$ jump in PA occurs, while the basic morphology
of the emission beam does not change.

\begin{figure}
\centerline{\psfig{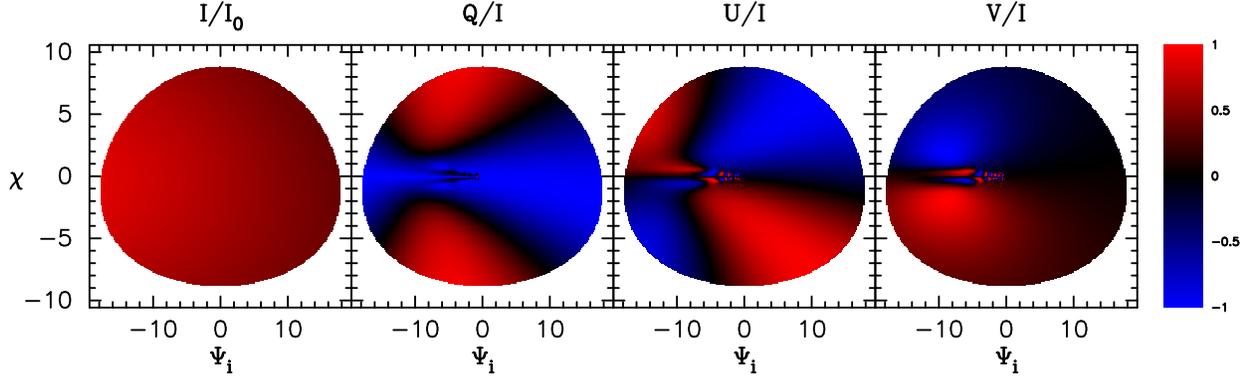}}
\caption{Two-dimensional polarization map of pulsar emission beam. The
  four panels correspond to the four Stokes parameters $I$, $Q/I$,
  $U/I$, $V/I$. The filled region is the emission beam from the open
  field region.  The values of the four Stokes parameters are shown as
  different colors (with red for positive value while blue for
  negative). The map is obtained by computing the observed wave
  polarization for different impact angle $\chi$ (which varies from
  $-\theta_{\rm beam}$ to $\theta_{\rm beam}$, here $\theta_{\rm
    beam}\simeq 8.8^o$) and the rotation phase $\Psi_i$.  The other
  (fixed) parameters are: surface magnetic field $B_\ast=10^{12}$\,G,
  NS period $P=1$\,s, wave frequency $\nu=1$\,GHz, plasma density
  $\eta=N/N_{\rm GJ}=400$ ($N=N_e+N_p$ and $N_e=N_p$), Lorentz factor
  $\gamma=100$ (with $\Delta\gamma/\gamma\simeq 0$), inclination angle
  $\alpha=30^o$, and emission height $r_{\rm
    em}=50R_\ast$. }\label{fig:2D_1}
\end{figure}

\begin{figure}
\centerline{\psfig{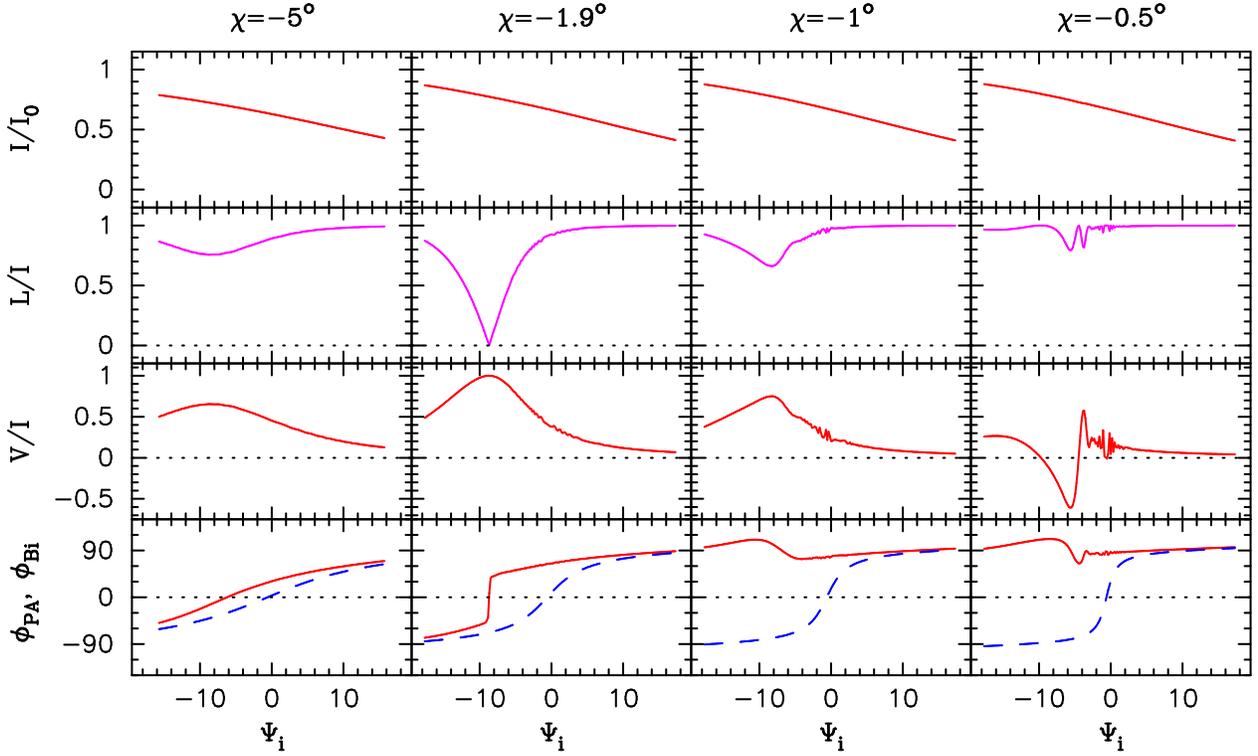}}
\caption{Intensity and polarization profiles of pulsar emission
  beam. The four columns correspond to four fixed impact angles
  $\chi=-5^o$, $-1.9^o$, $-1^o$ and $-0.5^o$, respectively, which are
  four sections in Fig.~\ref{fig:2D_1}.  In each column, we plot the
  Stokes parameters $I/I_0$ (top panel), $L/I$ (second panel), $V/I$
  (third panel), $\phi_{\rm PA}$ and $\phi_{B\rm i}$ (bottom panel,
  the solid line for $\phi_{\rm PA}$ and the dashed for $\phi_{B\rm
    i}$).  Note that there exists a $90^o$ jump of $\phi_{\rm PA}$
  near $\Psi_{\rm i}= -9^o$ in the bottom panel of the $\chi=-1.9^o$
  column.  The other parameters are the same as in
  Fig.~\ref{fig:2D_1}.}\label{fig:2D_4t}
\end{figure}

\section{Conclusion and Discussion}

We have studied the evolution of radio emission polarization in a
rotating pulsar magnetosphere filled with relativistic streaming pair
plasma. We quantify and compare the relative importance of several key
propagation effects that can influence the observed radio polarization
signals, including wave mode coupling, cyclotron absorption,
propagation through quasi-tangential (QT) region, and mode
circularization (due to asymmetric distributions of electrons and
positrons). We use numerical integration of the photon polarization
along the ray to incorporate all these propagation effects
self-consistently within a single framework. We find that, for typical
parameters of the magnetospheric plasma produced by pair cascade, and
for an initially $100\%$ linear polarized radio wave, the final
intensity and polarization position angle are modified by the
propagation effects, and significant circular polarization can be
generated.

We find that the most important propagation effects are cyclotron
absorption, wave mode coupling and QT effect. Generally, cyclotron
absorption occurs after the wave mode coupling [$r_{\rm cyc}>r_{\rm
    pl}$, see eq.~(\ref{eq:r_pl2r_cyc})]. Thus, it only changes the
total wave intensity and does not modify the wave polarization
($\phi_{\rm PA}$, $V/I$). For a large impact angle $|\chi|$ and/or
relatively low plasma density, the final wave polarization angle
$\phi_{\rm PA}$ is determined by the azimuthal angle of $\vecB$ field
at the polarization limiting radius $r_{\rm pl}$, and the observed
circular polarization is determined by the value of $\Lambda=r_{\rm
  pl}\phi_B'$ at $r=r_{\rm pl}$ [see eqs.~(\ref{eq:Lambda}) --
  (\ref{eq:VI})].  In this case, the observed $\phi_{\rm PA}$ profile
is similar to the prediction of the Rotating-Vector Model (RVM),
except for a phase shift by the amount $r_{\rm pl}/r_{\rm lc}$ [see
  eq.~(\ref{eq:PAshift})]; the circular polarization has a single sign
across the emission beam (see Fig.~\ref{fig:1D_diffeta}).  For a small
impact angle and/or high plasma density, the QT effect becomes
important, the final polarization profiles are more irregular: a
$90^o$ sudden jump in PA may occur at certain phase, accompanied by
large circular polarization.  For very small $|\chi|$, the circular
polarization may change signs for at different phases (see the right
column of Fig.~\ref{fig:2D_4t}).

In this paper, we have adopted the simplest (and minimum) assumptions
about the property of the magnetospheric plasma and the intrinsic
radio emission mechanism (see below). Nevertheless, our results
already show great promise in explaining a number of otherwise
puzzling observations: \\
(i) It has been observed that in some single-pulse pulsars, the
intensity profile deviates from the single guassion shape.  One
possible reason is that cyclotron absorption depends on the rotation
phase (because the ray passes through different region of the
magnetosphere), as discussed in section~5.2. Thus, even when the
initial intensity profile from the emission beam is a Gaussion, the
observed profile can be non-gaussion.  \\
(ii) For the so-called conal double type pulsars, which in our model
corresponds to large impact angle $\chi$, the relationship between the
single sign of the circular polarization and the derivation of
$\phi_{\rm PA}$ (see Han et al. 1998) can be easily understood by the
wave mode coupling effect. According to eqs.~(\ref{eq:GG_n3}) and
(\ref{eq:PAsign}), an increasing $\phi_{\rm PA}$ corresponds to the
left-hand circular polarization ($V>0$) while a decreasing $\phi_{\rm
  PA}$ corresponds to the right-hand ($V<0$) one \footnote{Here we
  define the circular polarization as seem from the receiver, which is
  different from the defination in electrical engineering used by Han
  et al. (1998)}.  \\
(iii) According to our calculation, there exists a special impact
angle $\chi_{\rm jump}$, where the observed $\phi_{\rm PA}$ profile
has a $90^o$ jump (orthogonal polarization mode) and this is
accompanied by the maximum circular polarization. This feature may be
helpful to explain the polarization profile of PSR J1920+2650
(Figure~\ref{fig:J1920+2650}; see Han et al. 2009).  (iv) For a very
small impact angle, which corresponds to the core emission, the QT
effect can cause the sign reversal of circular polarization, which is
observed in the core components of many pulsars (e.g. Radhakrishnan \&
Rankin 1990; Han et al. 1998; You et al. 2006).

Our calculations in this paper have relied on several simplifying
assumptions.  For example, we have assumed that the radio emission is
from the same height for different rotation phases, that the density
parameter ($\eta=N/N_{\rm GJ}$) of the magnetospheric plasma is
constant everywhere in the emission cone and along the photon
trajectory, and that the plasma electrons and positrons have the same
for bulk Lorentz factors.  In future works, we plan to consider models
with varying emission heights, as well as non-trivial
electron/positron spatial and velocity distributions.  We did not
include the small but finite emission cone (angle $1/\gamma$) in our
model, and assumed that the initial polarization of photon is always
O-mode for different rotation phases.  However, different emission
mechanisms could give different initial polarization states. We will
also be interested in studying the propagation effects on the
individual/subpluse emissions, since they may more directly reflect
the underlying radio emission processes.

\begin{figure}
\centerline{\psfig{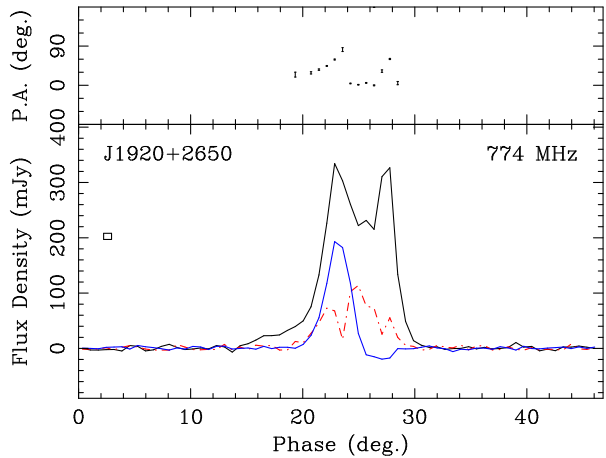}}
\caption{Intensity and polarization profiles of PSR J1920+2650. In the
  lower panel, the upper thick line is the total intensity ($I$), the
  dash-dot-dashed line is the linear polarized intensity $L$, and the
  thin line is the circularly polarized intensity $V$. This pulsar
  shows extremely strong circular polarization with $V/I\simeq 64\%$
  in its first major component.  }\label{fig:J1920+2650}
\end{figure}

\section*{Acknowledgments}

This work has been supported in part by NASA Grant NNX07AG81G and NSF
grants AST 0707628. Authors are also supported by the National Natural
Science Foundation of China (10773016, 10821001 and 10833003) and the
Initialization Fund for President Award winner of Chinese Academy of
Sciences.


\appendix

\newpage

\label{lastpage}


\begin{thebibliography}{longestkeymustbeshorterthanthis99}


\bibitem[]{} Allen, M. C. \& Melrose, D. B. 1982, PASAu, 4, 365

\bibitem[]{} Arons, J. \& Barnard, J. J. 1986, ApJ, 302, 120

\bibitem[]{} Backer, D. C.; Rankin, J. M.; Campbell, D. B. 1976, Nature, 263, 202

\bibitem[]{} Barnard, J. J. \& Arons, J. 1986, ApJ, 302, 138

\bibitem[]{} Beloborodov, A. M. \& Thompson, C. 2007, ApJ, 657, 967

\bibitem[]{} Blandford, R. D. \& Scharlemann, E. T. 1976, MNRAS, 174, 59

\bibitem[]{} Blaskiewicz, M., Cordes, J. M. \& Wasserman, I. 1991, ApJ, 370, 643

\bibitem[]{} Cheng, A. F. \& Ruderman, M. A. 1979, ApJ, 229, 348

\bibitem[]{} Cognard, I.; Shrauner, J. A.; Taylor, J. H. \& Thorsett, S. E. 1996, ApJ, 457, 81

\bibitem[]{} Cordes, J. M. 1978, ApJ, 222, 1006

\bibitem[]{} Cordes, J. M.; Rankin, J. \& Backer, D. C. 1978, ApJ, 223, 961

\bibitem[]{} Daugherty, J. K. \& Harding, A. K. 1982, ApJ, 252, 337

\bibitem[]{} Fussell, D., Luo, Q. \& Melrose, D. B. 2003, MNRAS, 343, 1248

\bibitem[]{} Gould, D. M. \& Lyne, A. G. 1998, MNRAS, 301, 235

\bibitem[]{} Han, J. L., Demorest, P. B., van Straten, W. \& Lyne, A. G. 2009, ApJS, 181, 557

\bibitem[]{} Han, J. L.; Manchester, R. N.; Xu, R. X. \& Qiao, G. J. 1998, MNRAS, 300, 373

\bibitem[]{} Hibschman, J. A. \& Arons, J. 2001, ApJ, 554, 624

\bibitem[]{} Johnston, S., Ball, L., Wang, N. \& Manchester, R. N. 2005, MNRAS, 358, 1069

\bibitem[]{} Kazbegi, A. Z., Machabeli, G. Z. \& Melikidze, G. I. 1991, MNRAS, 253, 377

\bibitem[]{} Kijak, J. \& Gil, J. 2003, A\&A, 397, 969

\bibitem[]{} Kramer, M., Xilouris, K. M., Jessner, A., Lorimer, D. R., Wielebinski, R.\& Lyne, A. G., 1997, A\&A, 322, 846

\bibitem[]{} Lai, D. \& Ho, W. C. G. 2003, ApJ, 588, 962

\bibitem[]{} Lai, D. \& Ho, W. C. G. 2002, ApJ, 566, 373

\bibitem[]{} Luo, Q. H. \& Melrose, D. B., 2004, in Camilo F., Gaensler B. M., eds, Proc. IAU Symp. 218, Young Neutron Stars and Their Environments. Astron. Soc. Pac., San Francisco, p. 381

\bibitem[]{} Luo, Q. H. \& Melrose, D. B. 2001, MNRAS, 325, 187

\bibitem[]{} Lyne, A. G. \& Manchester, R. N. 1988, MNRAS, 234, 477

\bibitem[]{} Lyubarsky, Y. 2008, in 40 YEARS OF PULSARS, eds. C. Bassa, Z. Wang, A. Cumming, \& V. M. Kaspi, AIP Conf. Ser., 983, 29

\bibitem[]{} Lyubarskii, Y. E. \& Petrova, S. A., 1998, Ap\&SS, 262, 379

\bibitem[]{} Manchester, R. N.; Taylor, J. H.; Huguenin, G. R. 1975, ApJ, 196, 83

\bibitem[]{} McKinnon, Mark M. \& Stinebring, Daniel R. 2000, ApJ, 529, 435

\bibitem[]{} Medin, Z. \& Lai, D. 2009, submitted to MNRAS

\bibitem[]{} Melrose, D. B. 2003, in Radio Pulsars, eds. M. Bailes, D. J. Nice, \& S. E. Thorsett, AIP Conf. Ser., 302, 179

\bibitem[]{} Melrose, D. B. 1979, AuJPh, 32, 61

\bibitem[]{} Mitra, D. \& Li, X. H. 2004, A\&A, 421, 215

\bibitem[]{} Petrova, S. A. 2006, MNRAS, 366, 1539

\bibitem[]{} Petrova, S. A. \& Lyubarskii, Y. E. 2000, A\&A, 355, 1168

\bibitem[]{} Radhakrishnan, V. \& Cooke, D. J. 1969, ApJ, 3, 225

\bibitem[]{} Radhakrishnan, V. \& Rankin, J. M. 1990, ApJ, 352, 258

\bibitem[]{} Rafikov, R. R. \& Goldreich, P. 2005, ApJ, 631, 488

\bibitem[]{} Rankin, J. M. 1983, ApJ, 274, 333

\bibitem[]{} Stinebring, D. R., Cordes, J. M., Rankin, J. M., Weisberg, J. M., \& Boriakoff, V. 1984a, ApJS, 55, 247

\bibitem[]{} Stinebring, D. R., Cordes, J. M., Weisberg, J. M., Rankin, J. M., \& Boriakoff, V. 1984b, ApJS, 55, 279

\bibitem[]{} Thompson, C., Lyutikov, M., \& Kulkarni, S. R. 2002, ApJ, 574, 332

\bibitem[]{} van Adelsberg, M. \& Lai, D. 2006, MNRAS, 373, 1495

\bibitem[]{} van Hoensbroech, A., Lesch, H. \& Kunzl, T. 1998, A\&, 336, 209

\bibitem[]{} Wang, C. \& Lai, D. 2007, MNRAS, 377, 1095

\bibitem[]{} Wang, C. \& Lai, D. 2009, MNRAS, 398, 515

\bibitem[]{} Weisberg, J. M., Cordes, J. M., Kuan, B., Devine, K. E., Green, J. T. \& Backer, D. C. 2004, ApJS, 150, 317

\bibitem[]{} Weisberg, J. M., Cordes, J. M., Lundgren, S. C., Dawson, B. R., Despotes, J. T., Morgan, J. J., Weitz, K. A., Zink, E. C. \& Backer, D. C. 1999, ApJS, 121, 171

\bibitem[]{} Xilouris, K. M.; Seiradakis, J. H.; Gil, J.; Sieber, W. \& Wielebinski, R. 1995, A\&A, 293, 153

\bibitem[]{} Xu, R. X., Liu, J. F., Han, J. L. \& Qiao, G. J. 2000, ApJ, 535, 354

\bibitem[]{} You, X. P. \& Han, J. L. 2006, ChJAA, 6, 237

\end{thebibliography}
\end{document}